\documentclass[journal]{IEEEtran}

\usepackage{times,amsmath,amsfonts,amssymb,subfigure,xspace,lscape,epstopdf,amsthm,bm} 
\usepackage{mdframed,multirow,hyperref}
\usepackage{multicol}
\usepackage[all]{xy}
\usepackage{xcolor}
\usepackage[vlined,english,ruled,linesnumbered]{algorithm2e}
\SetKwComment{Comment}{$\triangleright$\ }{}

\newif\ifpdf
\ifx\pdfoutput\undefined
   \pdffalse
\else
   \pdfoutput=1
   \pdftrue
\fi
\ifpdf
   \usepackage{graphicx}
   \usepackage{epstopdf}
\else
   \usepackage{graphicx}
\fi

\usepackage{color}

\DeclareMathOperator*{\argmax}{arg\,max}

\usepackage{changepage}
\usepackage{hyperref}

\theoremstyle{plain}

\theoremstyle{definition}

\newtheorem*{defn*}{Definition}

\theoremstyle{remark}

\begin{document}

\title{Learning to Search for MIMO Detection}
\date{}

\author{Jianyong~Sun~\IEEEmembership{Senior~Member, ~IEEE}, Yiqing Zhang, Jiang Xue~\IEEEmembership{Senior~Member, ~IEEE} and Zongben Xu~\thanks{Corresponding author: Jiang Xue. Jianyong Sun,
Yiqing Zhang, Jiang Xue and Zongben Xu are with the School of Mathematics and Statistics, Xi'an Jiaotong University, Xi'an, China (e-mail: jy.sun@xjtu.edu.cn, zhangyiqing@stu.xjtu.edu.cn, x.jiang@xjtu.edu.cn, zbxu@xjtu.edu.cn).} }


\maketitle


\begin{abstract}
This paper proposes a novel learning to learn method, called learning to learn iterative search algorithm (LISA), for signal detection in a multi-input multi-output (MIMO) system. The idea is to regard the signal detection problem as a decision making problem over tree. The goal is to learn the optimal decision policy. In LISA, deep neural networks are used as parameterized policy function. Through training, optimal parameters of the neural networks are learned and thus optimal policy can be approximated. Different neural network-based architectures are used for fixed and varying channel models, respectively. LISA provides soft decisions and does not require any information about the additive white Gaussian noise. Simulation results show that LISA 1) obtains near maximum likelihood detection performance in both fixed and varying channel models under QPSK modulation; 2) achieves significantly better bit error rate (BER) performance than classical detectors and recently proposed deep/machine learning based detectors at various modulations and signal to noise (SNR) ratios both under i.i.d and correlated Rayleigh fading channels in the simulation experiments; 3) is robust to MIMO detection problems with imperfect channel state information; and 4) generalizes very well against channel correlation and SNRs.
\end{abstract}


\section{Introduction}

Multi-input Multi-output (MIMO) is a key technology in the fifth-generation (5G) wireless communication system. It is mostly used for improving the spectrum efficiency and channel capacity. In a MIMO system, several transmitting and receiving antennas are simultaneously used at the transmitter and the receiver end. In comparison with traditional Single-input Single-output (SISO) system, the MIMO system can make full use of space resources and increase the channel capacity without increasing the bandwidth since each antenna at the receiver can receive signals transmitted from all the transmitting antennas simultaneously. MIMO has been widely applied in various wireless communication systems due to its connection reliability and high transmission rate.

For MIMO, precoding, channel estimation and signal detection are key technologies for the 5G wireless communication system. Specifically, precoding is applied to overcome the interference between users and enhance the signal power received by each user. Channel estimation aims to provide the channel state information (CSI) to the transceiver. Signal detection is mostly used to recover the transmitted signal based on the received signals and CSI. This paper focuses on signal detection which is known to be NP-hard. It is of great challenge to recover the true signals after MIMO transmission, due to noise and inter-symbol interference.

\subsection{Related Work}

Many algorithms have been proposed to address the MIMO detection problem since 1960s~\cite{7244171}. Among these algorithms, maximum likelihood detection (MLD)~\cite{1054829} is able to find the global optimal solution as it searches over all possible transmitted signals exhaustively. Its time complexity is thus prohibitively high. Therefore, it has little practical use, but can be used as a baseline for performance measurement.

To reduce the computational complexity, some linear detection algorithms, such as matched filter (MF)~\cite{7244171}, zero-forcing (ZF)~\cite{4815548},  ZF with decision feedback (ZF-DF)~\cite{zf-df}, minimum mean square error (MMSE)~\cite{7244171} detector, and many others, have been developed. It is acknowledged that the accuracy of these detectors tend to be poor~\cite{7244171}.

Fortunately, for massive MIMO systems (i.e. MIMO with a high number of antennas), it has been proved that linear detectors can achieve near-optimal performance in debt to the channel harden phenomenon~\cite{6375940}. However, ZF and MMSE both require high-order matrix inversion. It is known that inversion can be very high in complexity and computationally unstable for high-order matrix. To reduce complexity and increase stability, two categories of methods have been proposed. First, matrix inversion can be approximated by taking the first few terms of its expansion in series, e.g. the Neumann series expansion (NSE)~\cite{7248580,7401703}. However, as the number of the truncated terms decreases, the complexity will be greatly increased. In~\cite{7370771,8638827}, the Neumann iteration method has been proposed to approximate matrix inversion iteratively. Although it promises to have a faster convergence than NSE, its performance is sensitive to initialization and the complexity is also high. The second type is to convert the matrix inversion problem into solving a linear equation. Any quick linear equation solving methods, such as Richardson~\cite{6966041}, Jocabi~\cite{7342925}, Gaussian-Seidel~\cite{6954512}, Successive over Relaxation (SOR) methods~\cite{7037314}\cite{7399337}, and others can be applied to recover the transmitted signal. In comparison with NSE, the complexity of these algorithms is usually reduced by an order of magnitude while maintaining a good detection performance.

In addition, a variety of nonlinear detection algorithms have been proposed in which compromises between computation complexity and detection accuracy are made. Representative algorithms include sphere decoding (SD)~\cite{1019833}\cite{Damen:2003:MDS:2263399.2271012}, semidefinite relaxation (SDR)~\cite{4475373}, approximate message propagation (AMP)~\cite{7282651}\cite{6778065} and so on. While having lower complexity than the MLD, these algorithms can achieve suboptimal detection performance.

Owing to the development of advanced optimization algorithms and the fast-growing of computing power, machine learning has made great achievements~\cite{mlbook}. Machine learning algorithms have been widely applied to many research areas, such as global optimization~\cite{sun19a,sun19b,shi19}, system biology~\cite{sun12}, etc. and industrials, such as Apple Siri, etc, to name a few. They have also been successfully applied to the MIMO detection problem. For example, Huang et al.~\cite{8422211} proposed to convert the MIMO detection problem into a clustering problem, which is then solved by the expectation-maximization algorithm. Simulation experiments showed that this method can achieve near MLD performance when the channel is fixed and perfectly known. However, it is not applicable to varying channel and hence has very limited practical use. Elgabli et al.~\cite{8335641} reformulates the MIMO detection problem as a Lasso~\cite{lasso} problem. It is then optimized by a two-stage ADMM~\cite{admm} algorithm. Experiments showed that it behaves well compared to classical detectors.

In recent years, deep learning has achieved great success in various fields, such as computer vision, speech recognition, natural language processing, and so on~\cite{dlbook}. Not surprisingly, there are also some attempts to apply deep learning methods to the MIMO detection problem. These deep learning-based detectors basically fall into two categories depending on how deep learning techniques are used.

In the first category, deep learning techniques are used as feature extractor. Yan et al. proposed a detector called AE-ELM for the MIMO-OFDM system~\cite{7966042}, in which an auto-encoder (AE) network is combined with extreme learning machine (ELM) to recover the transmitted signal for the OFDM (Orthogonal Frequency Division Multiplexing) system. The AE is used to extract features of the received signal, and the features are classified by the ELM. Experiments showed that AE-ELM can achieve a detection performance similar to MLD in the MIMO-OFDM systems.

In deep learning, how to design the structure of a neural network is a key issue that needs to pay great attention. Recent development addresses this issue by proposing a learning to learn approach~\cite{10.1093/nsr/nwx099}\cite{lecun}. The main idea is to unfold an iterative algorithm (or a family of algorithm) to a fixed number of iterations. Each iteration is considered as a layer and the unfolded structure is called a deep neural network (DNN).  This model-driven deep learning approach can achieve or exceed the performance of corresponding iterative algorithms~\cite{10.1093/nsr/nwx099}~\cite{lecun} since the advantages of model-driven and data-driven approach are effectively complementary to each other.

Few learning to learn approaches have been developed for continuous and combinatorial optimization problems. Andrychowicz et al.~\cite{Andrychowicz2016Learning} proposed to learn the descent direction by long short term memory (LSTM)~\cite{Hochreiter1997Long} recurrent neural network (RNN) for continuous optimization algorithms with differentiable objective functions. In Li et al.~\cite{Li2016Learning}, the decision of the iterative change is formalized under the Markov decision process framework, in which a feed-forward neural network is used to model the policy function. Chen et al.~\cite{Chen2016Learning} proposed a learning to learn approach for black-box optimization problems, in which LSTM is used to model the iterative change. The learned algorithm compares favorably against Bayesian optimization~\cite{mockus2012bayesian} for hyper-parameter tuning. Dai et al.~\cite{dai2017} proposed to learn heuristics for combinatorial optimization problems, such as traveling salesman problem, maximum cut problem, and others. Experimental results showed that their algorithm performs well and can generalize to combinatorial optimization problems with different sizes.

Learning to learn approach has also been developed~\cite{8804165} for MIMO detection. For examples, Samuel et al.~\cite{DBLP:journals/corr/abs-1805-07631} proposed a detector called DetNet. It is the unfolding of a projected gradient descent algorithm. The iteration of the projected gradient descent algorithm is of the following form
\begin{equation}  \hat{\bm{s}}_{k+1} = {\cal P} \left[ \hat{\bm{s}}_k - \delta_k \bm{H}^\intercal \bm{y} + \delta_k  \bm{H}^\intercal  \bm{H}\hat{\bm{s}}_k\right]\label{detnet}\end{equation}where $\bm{H}$ is the channel matrix, $\bm{y}$ is the received signal, $\hat{s}_k$ is the signal estimate in the $k$-th iteration, $\delta_k$ is the step size,  and $\cal P$ is a non-linear projection operator. A neural network is designed to approximate the projection operator. Simulation experiments show that DetNet performs better than some traditional detectors. However, in terms of the detection accuracy, there is still a big gap between DetNet and MLD. Actually, there have no detectors in literature that are comparable with MLD in the varying channel scenario. In addition, DetNet requires that the number of receiving antennas is more than that of the transmitting antennas. This condition may not be always true in reality.

Gao et al.~ \cite{EasyChair:376} proposed to simplify the structure of DetNet by reducing the input, changing the connection structure from the fully connection to a sparse one and modifying the loss function. These simplifications reduce the complexity and improve the detection accuracy to some extent. Corlay et al.~\cite{DBLP:journals/corr/abs-1812-01571} proposed to change the sigmoid activation function used in DetNet to a multi-plateau version and used two networks with different initial values to detect the transmitted signals simultaneously. The detection performance can be improved by selecting the solution which has a smaller loss function. Similar to DetNet, OAMP-Net~\cite{DBLP:journals/corr/abs-1809-09336}~\cite{DBLP:journals/corr/abs-1907-09439} is designed by unfolding the orthogonal AMP (OAMP) algorithm. It is claimed that OAMP-Net requires a short training time, and is able to adapt to varying channels. However, OAMP-Net needs to estimate the noise variance in advance. Tan et al.~\cite{8646369} proposed to unfold a modified message passing detection (MDP) algorithm~\cite{6952193} for signal detection, in which the MDP parameters are learned by a DNN.

For large-scale overloaded MIMO channels (i.e. the number of the receiving antennas is less than the number of transmitting antennas), Imanishi et al.~\cite{2018arXiv180610827T}~\cite{DBLP:journals/corr/abs-1812-10044} proposed a trainable iterative detector based on the iterative soft thresholding algorithm (ISTA). In their algorithm, the sensing matrix in ISTA is replaced by the pseudo-inverse of the channel matrix. In low-order modulation, the detector can achieve comparable detection performance to the iterative weighted sum-of-absolute value optimization algorithm~\cite{7760475} with low complexity. However, the performance of the detector deteriorates seriously when the modulation order becomes higher. Tan et al.~\cite{Tan2018Improving} proposed to use DNN to improve the performance of the belief propagation (BP) algorithm~\cite{7345035} for MIMO detection. Their methods, named DNN-BP and DNN-MS, are the unfolding of two modified BP detectors, respectively. It was claimed that DNN-BP and DNN-MS both improve the detection accuracy of the BP algorithm. Wei et al.~\cite{DBLP:journals/corr/abs-1906-03814} and Shirkoohi et al.~\cite{Shirkoohi2019AdaptiveNS} proposed to design the structure of the detection network by unfolding the conjugate gradient method and ISTA, respectively, which have achieved promising results for massive MIMO channels.

Although those learning to learn approaches have achieved substantial improvement over traditional detectors, there is still a big gap between these detectors and MLD.

\subsection{Main contributions}

In this paper, we propose a novel learning to learn method, named learning to learn iterative search algorithm (LISA), for MIMO detection. Different from the present learning to learn approaches, we do not unfold existing iterative detection algorithm to a deep network. Rather, we first propose to iteratively construct a solution to the detection problem by taking deep neural network as building block. The constructive algorithm is then unfolded into a shallow network. In the construction process, the symbols of the transmitted signal are recovered one by one. On detecting each symbol, soft decisions are provided to meet the requirements of communication systems.

Experimental results show that LISA has a strong ability of generalization. Once trained, LISA can adapt to different models and signal to noise ratios (SNRs). It also performs well for both fixed and varying channels. Particularly, experimental results show that LISA performs significantly better than existing learning to learn approaches for fixed and varying channel models at different modulation and SNRs. Surprisingly, LISA can achieve near-MLD performance in both fixed and varying channel scenarios in the QPSK modulation.

The rest of the paper is organized as follows. In Section~\ref{problem}, the MIMO detection problem is introduced. Section~\ref{algorithm} presents LISA. Experimental results are given in Section~\ref{experiments}. Section~\ref{conclusion} concludes the paper.


\section{The Problem}\label{problem}

In this section, we describe the MIMO model and review the reformulation of the MIMO detection problem based on QL-decomposition.

\subsection{Notations}

Throughout the paper, we use lowercase letter to denote scalar (e.g. $s$), boldface lowercase letter to denote vector (e.g. $\bm{s}$), and boldface uppercase letter to denote matrix (e.g. $\bm{H}$). ${\bm{H}}^\intercal$ denotes the transpose of the matrix $\bm{H}$. $\mathcal{CN}({\bm{\mu}},\sigma^2\bm{I})$ denotes the complex Gaussian distribution with mean $\bm{\mu}$ and covariance matrix $\sigma^2\bm{I}$. $Re(\cdot)$ and $Im(\cdot)$ denote the real and imaginary part of a complex number, respectively.

\subsection{The MIMO detection problem}

Consider a MIMO system with $N_T$ transmitting antennas at the transmitter and $N_R$ receiving antennas at the receiver. Assume that at some time step $t$, the transmitted symbol vector at the transmitter is $\bar{\bm{s}}=\left(\bar{s}_1,\bar{s}_2,\cdots,\bar{s}_{N_T}\right) \in \bar{\mathcal{A}}^{N_T}$, where $\bar{s}_i \in \bar{\mathcal{A}}, i=1,2,\cdots,N_T$ represents the transmitted symbol from the $i$-th transmitting antenna, and $\bar{\mathcal{A}}$ is a finite alphabet related to the constellation.

The MIMO detection problem is to recover $\bar{\bm{s}}$ from the received symbol vector $\bar{\bm{y}}=\left(\bar{y}_1,\bar{y}_2, \cdots,\bar{y}_{N_R}\right)\in \mathcal{C}^{N_R}$ observed at the receiver. Here $\bar{\bm{y}}$ can be described as follows:
\begin{equation}\label{origpro}
 \bar{\bm{y}}=\bar{\bm{H}}\bar{\bm{s}}+\bar{\bm{n}}
\end{equation} where $\bar{\bm{H}}\in\mathcal{C}^{{N_R}\times{N_T}}$ represents the complex channel matrix, each element
$\bar{h}_{ij}$ is the path gain from the $j$th transmitting antenna to the $i$th receiving antenna.
The value of $\bar{h}_{ij}$ is drawn from the i.i.d. complex Gaussian distribution with zero mean and unit variance. $\bar{\bm{n}} \in \mathcal{C}^{N_R}$ represents the additive white Gaussian noise (AWGN) at the receiver, i.e. \begin{equation}\bar{\bm{n}} \sim \mathcal{CN}\left(0,\sigma^2 \bm{I}_{N_R}\right).\end{equation}

It is difficult to solve the complex-valued MIMO channel model directly. Without loss of generality, we use an equivalent real-valued channel model by separating the real and imaginary parts of the transmitted and received symbol vectors. The following equations show how to convert the complex-valued model to a real-valued one,
\begin{align*}
  \bm{y}&=\left[\begin{array}{c}
  Re(\bar{\bm{y}})\\
  Im(\bar{\bm{y}})
  \end{array}
  \right],\
  \bm{H}=\left[\begin{array}{cc}
  Re(\bar{\bm{H}}) & -Im(\bar{\bm{H}})\\
  Im(\bar{\bm{H}}) & Re(\bar{\bm{H}})
  \end{array}
  \right], \\
  \bm{s}&=\left[\begin{array}{c}
  Re(\bar{\bm{s}})\\
  Im(\bar{\bm{s}})
  \end{array}
  \right],\ \ \
  \bm{n}=\left[\begin{array}{cc}
  Re(\bar{\bm{n}})\\
  Im(\bar{\bm{n}})
  \end{array}
  \right].
\end{align*}

In this way, the MIMO channel model can be written as:
\begin{equation}\label{problemx}
  \bm{y}=\bm{Hs}+\bm{n}
\end{equation}where ${\bm{y}} \in R^{2N_R}$, ${\bm{H}} \in R^{2N_R \times 2N_T}$, ${\bm{n}} \in R^{2N_R}$, ${\bm{s}} \in \mathcal{A}^{2N_T}$, and $\mathcal{A}=Re(\mathcal{\bar{A}})$, i.e. the elements in $\mathcal{A}$ is the real part of elements in $\mathcal{\bar{A}}$.  We assume the size of $\mathcal{A}$ is $M$. Its value depends on the modulation mode.

In the MIMO detection problem, we assume perfect channel state information. The goal is to recover the transmitted symbol vector $\bm{s}$ as accurately as possible when we observe the received signal $\bm{y}$.

The best detection algorithm to solve the MIMO detection problem is the maximum likelihood detector (MLD), which solves the problem based on the maximum likelihood criterion:
\begin{equation}\label{mlc}
  \min\limits_{{\bm{s}}\in\mathcal{A}^{2N_T}}||\bm{y}-\bm{H}\bm{s}||_2^2
\end{equation}

The MLD searches all possible solutions in the solution space and selects the one that minimizes the term in Eq.~(\ref{mlc}) as the transmitted signal. It has a prohibitively high time complexity (exponential in the number of $N_T$). Apparently, the MLD is not practicable for system with large $N_T$. We have found that no existing detectors are able to compare with the MLD in terms of detection accuracy.

\subsection{QL decomposition}

Our constructive algorithm is built upon the transformation of the detection problem through QL decomposition. Consider the QL-decomposition of the channel matrix $\bm{H}$, i.e., $\bm{H}=\bm{Q}\bm{L}$, where ${\bm{Q}} \in R^{ 2N_R \times 2N_T} $ is an orthogonal matrix, and ${\bm{L}} \in R^{2N_R \times 2N_T}$ is a lower triangular matrix. Then Problem~(\ref{mlc}) can be converted to its equivalent form:
\begin{equation}
 \min\limits_{{\bm{s}}\in\mathcal{A}^{2\emph{N}_T}}||\bm{\widetilde{y}}-\bm{L}\bm{s}||_2^2\label{ql}
\end{equation}where $\widetilde{\bm{y}}=\bm{Q}^T\bm{y}$. Expanding the $l_2$-norm in Eq.~(\ref{ql}), we get
the following equivalent form:
\begin{equation}
\min\limits_{\bm{s}\in\mathcal{A}^{2\emph{N}_T}}\{f_1(s_1)+f_2(s_1,s_2)+\cdots+f_{2\emph{N}_T}(s_1,\cdots,s_{2\emph{N}_T})\}\label{qlform}
\end{equation}where
\begin{eqnarray}
f_{k}(s_1,\cdots,s_k)=\left(\widetilde{y}_k-\sum_{l=1}^{k}l_{k,l}s_l\right)^2
\end{eqnarray}and $l_{k,l}$ is the element of $\bm L$ at $(k,l)$.

Problem~(\ref{qlform}) shows that we can detect the symbol from $s_1$ to $s_{2N_T}$ one by one recursively. That is to say, we can recover $s_1$ first,  calculate $f_1(s_1)$; recover $s_2$ based on the known $s_1$ and calculate $f_2(s_1,s_2)$, repeat this process until $s_{2N_T}$ is recovered. Along the search procedure, we need to compare totally $|\mathcal{A}|^{2N_T}$ possible solutions, and the one that minimizes $f_1(s_1)+f_2(s_1,s_2)+\cdots+f_{2\emph{N}_{T}}(s_1, \cdots,s_{2\emph{N}_T})$ is the optimal signal recovered. Many detectors, such as ZF~\cite{4815548}, ZF-DF~\cite{zf-df}, SD~\cite{1019833}\cite{Damen:2003:MDS:2263399.2271012} and SDR~\cite{4475373}, are built upon QL decomposition.

\section{Learning to do iterative search}


\subsection{The Decision Making Problem}

Note that Problem~(\ref{qlform}) can actually be visualized as a  decision tree with $N_T+1$ layers, $|\cal A|$ branches stemmed from each non-leaf node, and $|{\cal A}|^{N_T}$ leaf nodes. A cumulative metric $f_1(s_1) + f_2(s_1, s_2) + \cdots + f_k(s_1, s_2, \cdots, s_k)$ is associated at each node, and a branch metric $ f_k(s_1, s_2, \cdots, s_k)$ at each branch (except the root).

The detection of the transmitted signal can then be considered as a decision making problem. At each node, a decision is to make: which branch shall the detector choose to go? In ZF, ZF-DF, SD and the fixed-complexity sphere decoding (FCSD)~\cite{fcsd}, different decision strategies have been used. For examples, in ZF and ZF-DF, $s_k$ is estimated as:
\begin{equation}\hat{s}_k = \arg\min_{s_k \in {\cal A}} f_k (\hat{s}_1, \hat{s}_2, \cdots, \hat{s}_{k-1}, s_k)\label{zf}\end{equation}where $\hat{s}_1, \hat{s}_2, \cdots, \hat{s}_{k-1}$ are the estimated signals. That is, ZF-DF chooses the branch that minimize the cumulative metric. In the SD, at each node, it will skip the branches emanating from the node if it has a large radium $R$ such that $\|\widetilde{\bm{y}} - \bm{L}\bm{s}\|_2^2 > R$.

From a decision-tree perspective, ZF and ZF-DF searches just a single path down from the root. It is clearly not able to obtain a satisfactory performance. On the other hand, the MLD searches all the branches. It is clearly not economic. The SD tries to make a balance on the detection complexity and accuracy by setting a proper $R$. The value of $R$ determines how much branches the detector needs to search over. A smaller $R$ means to search over less branches, hence exhibits a lower complexity but a lower accuracy. In case $R = \infty$, the SD degenerates to the MLD.

In this paper, rather than using a fixed strategy as in the aforementioned detectors, we propose to use a deep neural network to adaptively make the decision. At each node, the decision is made based on the output of a deep neural network, while the information collected so far is used as input to the neural network. By this way, the detector does not need to search over the entire tree branches, but makes decisions purely based on the learned neural network. Apparently, this can accelerate the detection speed, but the accuracy depends entirely on the quality of the learned neural network.


\subsection{LISA}\label{algorithm}

In this section, we present the proposed learning to learn iterative search algorithm (LISA) for MIMO with fixed and varying channels, respectively. LISA consists of multiple blocks, and each block is a full solution construction procedure. In the following subsections, we present the procedure for fixed and varying channels, respectively.

Before presenting the procedure, let's consider the real-valued channel model in Eq.~(\ref{problemx}). It can be rewritten as follows when QL-decomposition is performed on $\bm{H}$.
\begin{equation}\label{model}
  \widetilde{\bm{y}}= \bm{Ls}+\widetilde{\bm{n}}
\end{equation}
where $\widetilde{\bm{y}}=\bm{Q}^\intercal\bm{y}, \widetilde{\bm{n}}=\bm{Q}^\intercal\bm{n}$.  Rewriting Eq.~(\ref{model}), the following formula can be established if the noise is ignored.
\begin{equation}\label{solver}
\begin{split}
\left\{
  \begin{aligned}
  \hat{s}_{1}=\,&\arg\min_{s_1 \in {\cal A}} f_{1}(\widetilde{y}_{1},l_{1,1})\\
  \hat{s}_{2}=\,& \arg\min_{s_2 \in {\cal A}} f_{2}(\widetilde{y}_{2},s_{1},l_{2,1},l_{2,2})\\
  &\cdots\\
  \hat{s}_{2\emph{N}_{T}}=\,& \arg\min_{s_{2N_t} \in {\cal A}} f_{2\emph{N}_{T}}(\widetilde{y}_{2\emph{N}_{T}},s_{1},s_{2},\cdots,s_{2\emph{N}_{T}-1},\\&l_{2\emph{N}_{T},1},l_{2\emph{N}_{T},2}\cdots,l_{2\emph{N}_{T},2\emph{N}_{T}})
  \end{aligned}
\right.
\end{split}
\end{equation}
From Eq.~(\ref{solver}), it is seen that the recovered symbol $\hat{s}_1$ is a function of $\widetilde{y}_1$ and $l_{1,1}$; $\hat{s}_2$ is a function of $\widetilde{y}_{2}, s_1, l_{2,1}$ and $l_{2,2}$, and so forth.

In traditional detectors such as ZF and ZF-DF, previously detected symbols are used in later detection. That is, the detection of $s_{{k}}$ depends on previously detected symbols $\hat{s}_1, \hat{s}_2, \cdots, \hat{s}_{{k}-1}$ (cf Eq.~(\ref{zf})). As mentioned before, this can reduce computational complexity but also detection accuracy. Further, solving functions ($f_1, \cdots, f_{2N_T}$) return solutions that are linear combinations of previously detected symbols and observations. This leads to exact solutions if there is no noise and inter-symbol interference. However, the solutions to problems in Eq.~(\ref{solver}) should be intrinsically non-linear to known information in practice due to the existence of noise and interference.



{To address this problem, we propose to use neural networks (which are non-linear function approximators) to approximate the solutions.

\subsubsection{The solution construction procedure for varying channel}

In this case, we assume that signals come from different channels. i.e., the channel is time-varying.

We propose to use LSTM~\cite{Hochreiter1997Long,6789445} to capture the non-linear relationship between the transmitted signal and observations.  LSTM is a known technique for processing sequential data in deep learning. It can be considered as a recursive function:
\begin{equation}\{C_{k}, h_{k}\} = \text{LSTM}(C_{{{k}}-1}, h_{{{k}}-1}, x_{{{k}}}; \theta_{{{k}}-1})\label{lstm}\end{equation}where $C_{{{k}}} (C_{{{k}}-1})$ is called the cell state which is for information flow along time, $C_{{{k}}}, h_{{{k}}}$ is the output at time ${{k}}$, $C_{{{k}}-1}$ and $h_{{{k}}-1}$ is the input at time ${{k}}$, and $x_{{{k}}}$ is the input (observation) at time ${{k}}$, respectively. $\theta_{{{k}}-1}$ is the parameter of LSTM. Usually $\theta_{{{k}}-1} = \theta$, i.e. the parameters of LSTM is shared along time. Fig.~\ref{lisa_varying1} shows the LSTM structure. As seen in the figure, LSTM has three gates, including forget gate $f_{{k}}$, input gate $i_{{k}}$ and output gate $o_{{k}}$, which are designed to remove or add information to the cell state $C_{{{k}}-1}$. Specifically, $f_{{k}}$ decides how much $C_{{{k}}-1}$ should be forgotten; $\tilde{C}_{{k}}$ represents the information extracted from the input $h_{{{k}}-1}$ and $x_{{k}}$; $i_{{k}}$ decides how much $\tilde{C}_{{k}}$ should be added to $C_{{{k}}-1}$; and finally, $o_{{k}}$ decides how much $C_{{k}}$ should be output. It is seen that there exists a high non-linearity between the input and output in LSTM. Please see the appendix for more details about LSTM.
\begin{figure}[htbp]
\small
\centering
\includegraphics[scale=0.4]{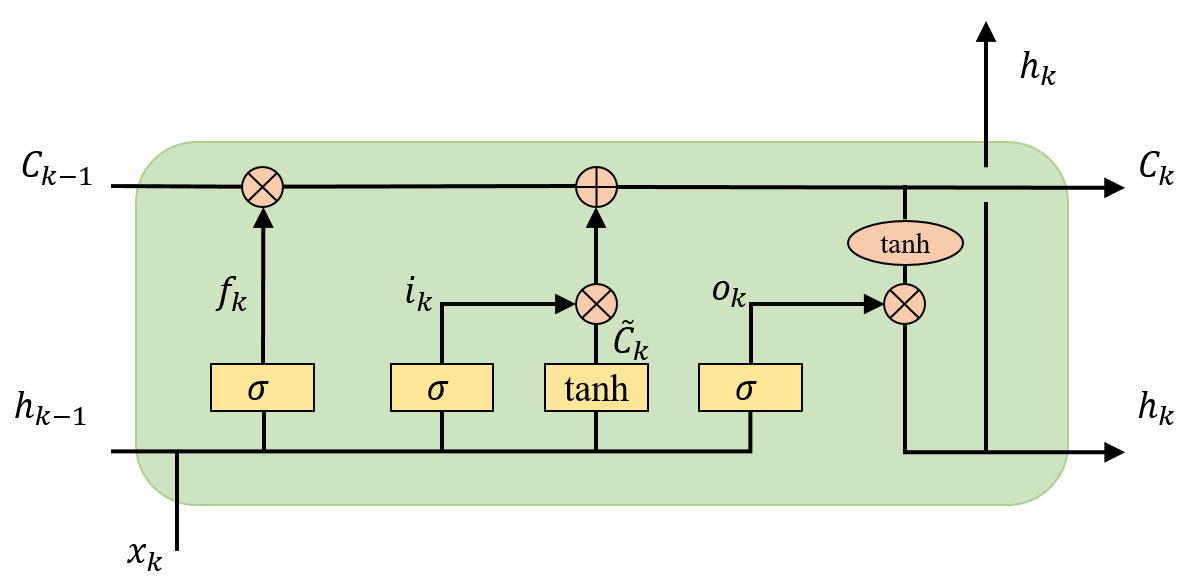}
\caption{The structure of LSTM.}\label{lisa_varying1}
\end{figure}

The basic structure of the solution construction procedure for varying channel is shown in Fig.~\ref{LISA_varying_channel}.
\begin{figure}[htbp]
\centering
\includegraphics[scale = 0.35]{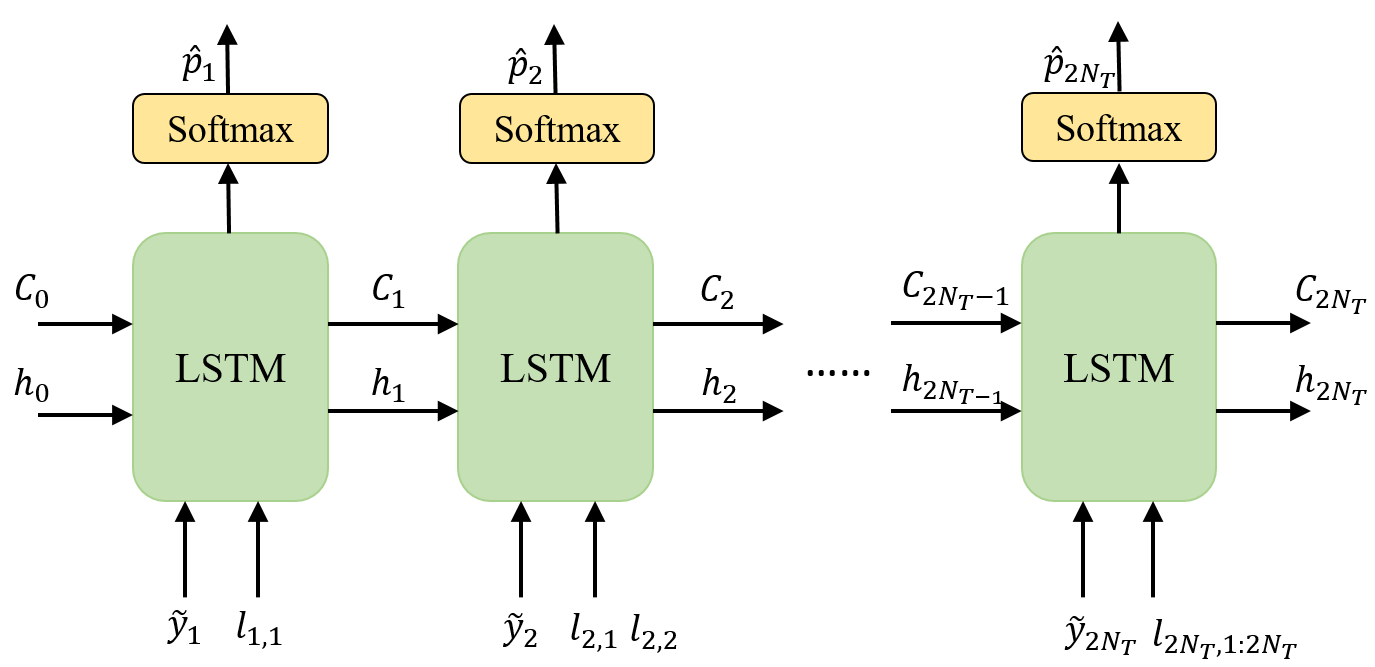}
\caption{The structure of the solution construction algorithm for varying channel.}
\label{LISA_varying_channel}
\end{figure}

From the figure, it is seen that when predicting $s_1$, the input to the LSTM is $x_1 = \{\widetilde{y}_1,l_{1,1}\}$ and $C_0, h_0$, while for $s_2$, the input is $x_2 = \{\widetilde{y}_{2},l_{2,1}, l_{2,2}\}$ and $C_1, h_1$, and so on. Generally, at each {step} ${{k}}$, given $x_{{k}}$ and previous memory $C_{{{k}}-1}$ and $h_{{{k}}-1}$, LTSM outputs $h_{{k}}$. In our approach, to make a soft-decision, a softmax layer is applied on the LSTM output $h_{{k}}$ at each signal recovery step. The softmax layer outputs the predicted probability distribution $\hat{\bm{p}}_{{k}} $ of $s_{{k}}$. Here $\hat{\bm{p}}_{{k}} = \left(\hat{p}_{{k}}^1, \cdots, \hat{p}_{{k}}^{M}\right)^\intercal $ is a probability vector, whose element represents the probability that $s_{{k}}$ equals to an element in $\cal A$. The estimated signal $\hat{s}_{{k}}$ is taken to be the constellation with the maximum probability of $\hat{\bm{p}}_{{k}}$. After obtaining $\hat{s}_{{k}}$, the observed information $x_{{k}}$ is then updated. The solution construction continues until the $2N_T$th symbol is detected.

Mathematically, for ${{k}} = 1, \cdots, 2N_T$, we have the following recursive functions:
\begin{eqnarray}
\{C_{{k}}, h_{{k}}\} &=& \text{LSTM}(C_{{{k}}-1}, h_{{{k}}-1}, x_{{{k}}}; \theta_{{{k}}-1}) \nonumber\\
\hat{\bm{p}}_{{k}} &=& \text{Softmax}(h_{{k}})\nonumber\\
\hat{s}_{{k}} &=& \arg\max(\hat{\bm{p}}_{{k}} )\nonumber\\
x_{{{k}}+1} &=&  \{\tilde{y}_{{{k}}+1}, l_{{{k}}+1,1:{{k}}+1}\} \nonumber
\end{eqnarray}where the softmax layer is a function of $h_{{k}}$ with parameters $\mathbf{w}_i, 1 \leq i \leq |\cal{A}|$, which can be stated as follows:\begin{equation}\hat{p}_{{k}}^i = \frac{\exp\left(h_{{k}}^\intercal \mathbf{w}_i \right)}{\sum_j \exp \left(h_{{k}}^\intercal \mathbf{w}_j \right)}.\end{equation}

It is seen that given all the information, including $\widetilde{\bm{y}}$ and $\bm{L}$, and parameters of the LSTMs, a signal vector can be recovered for Problem~(\ref{ql}) following Fig.~\ref{LISA_varying_channel}. Such a construction process is called a `block'.

\subsubsection{The solution construction procedure for fixed channel}

In this case, we assume that all signals come from a fixed channel which is perfectly known in advance. Compared with the varying channel scenario, this case is much easier to deal with. The basic structure of the solution construction procedure for fixed channel can thus be simplified.


There are two main changes in the fixed channel scenario in comparison with the varying case. First, when predicting $s_{{k}}$, the input is only $\widetilde{y}_{{k}}$, no $l_{{{k}},1:{{k}}}$. Values of the lower triangular matrix $\bm{L}$ are not required. Second, a fully connected DNN is used instead of LSTM. Except $C_{{k}}$, the rest are the same as those in the varying channel scenario. Please see Appendix for the structure of DNN.  The prediction for $s_{{k}}$ can be summarized as follows:
\begin{eqnarray}
h_{{k}} &=& \text{DNN}(h_{{{k}}-1}, x_{{{k}}};\theta_{{{k}}-1})\nonumber\\
\hat{\bm{p}}_{{k}} &=& \text{Softmax}(h_{{k}}) \nonumber\\
\hat{s}_{{k}} &=& \argmax(\hat{\bm{p}}_{{k}})\nonumber\\
x_{{{k}}+1} &=& \{\tilde{y}_{{{k}}+1}\} \nonumber
\end{eqnarray}

The basic structure for the fixed channel model is shown in Fig. \ref{LISA_fixed_channel}. Again the whole solution construction process is called a block.
\begin{figure}[htbp]
\small
\centering
\includegraphics[scale =0.32]{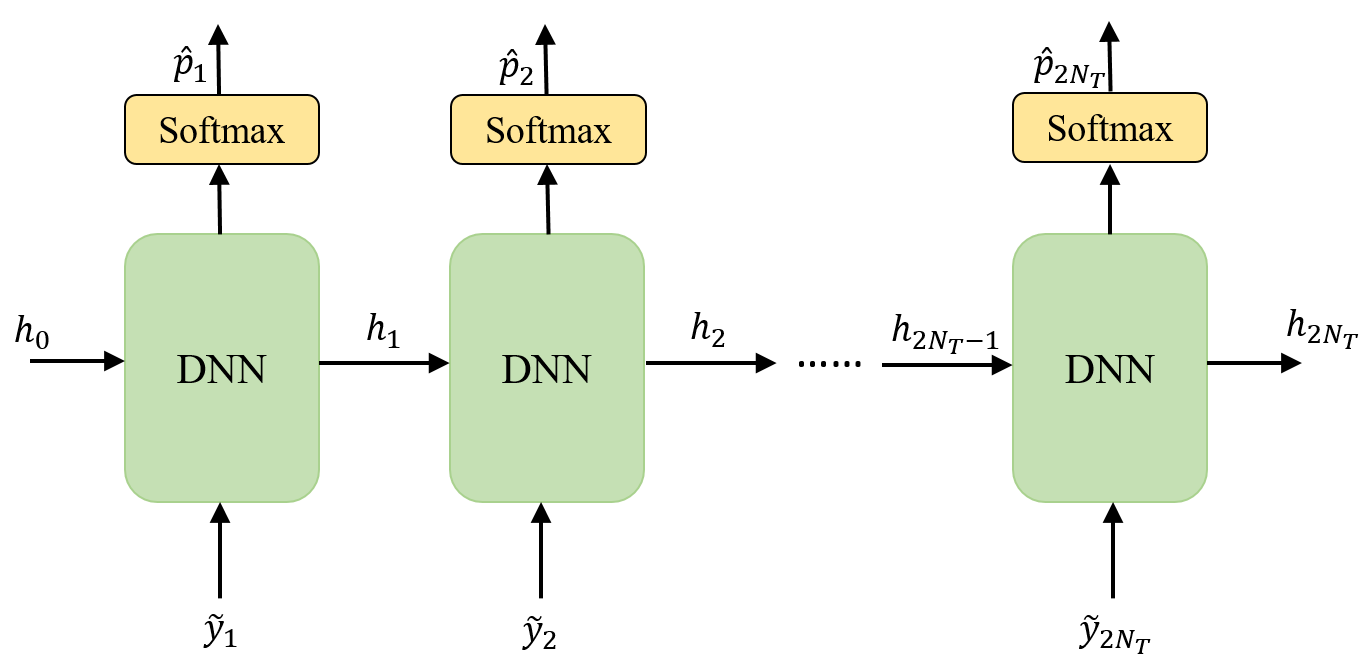}
\caption{The basic structure of the solution construction for fixed channel.}
\label{LISA_fixed_channel}
\end{figure}

\subsection{The architecture of LISA}

In a block, a transmitted symbol vector can be recovered one by one to reach a solution of Problem~(\ref{ql}). The obtained solution is not necessarily optimal or with high quality. To improve this solution, we propose to unfold the constructive algorithm to several blocks. The structure of LISA is shown in Fig.~\ref{RNN}.

 In the figure, $X$ represents the input, which contains the observation $\tilde{\bm{y}}$ and the low triangle matrix $\bm{L}$. It is the same for all the blocks. $\widehat{\bm{P}}_i = [\hat{\bm{p}}_1, \cdots, \hat{\bm{p}}_{2N_T} ]$ is the matrix of the probability obtained by the $i$-th block. $\bm{O}_{i}$ is the output at the $i$-th block. In the fixed channel scenario, the output is only $h_{2N_T}$, while in the varying channel scenario, the output includes $h_{2{N_T}}$ and $C_{2N_T}$.

Note that each block outputs not only a solution to the detection problem, but also some additional information. These additional information, including $C_{2N_T}$ and $h_{2N_T}$ for varying channel structure, and $h_{2N_T}$ for fixed channel structure, is helpful to improve the solution quality in the following blocks.
\begin{figure}[htbp]
\centering
\includegraphics[scale=0.34]{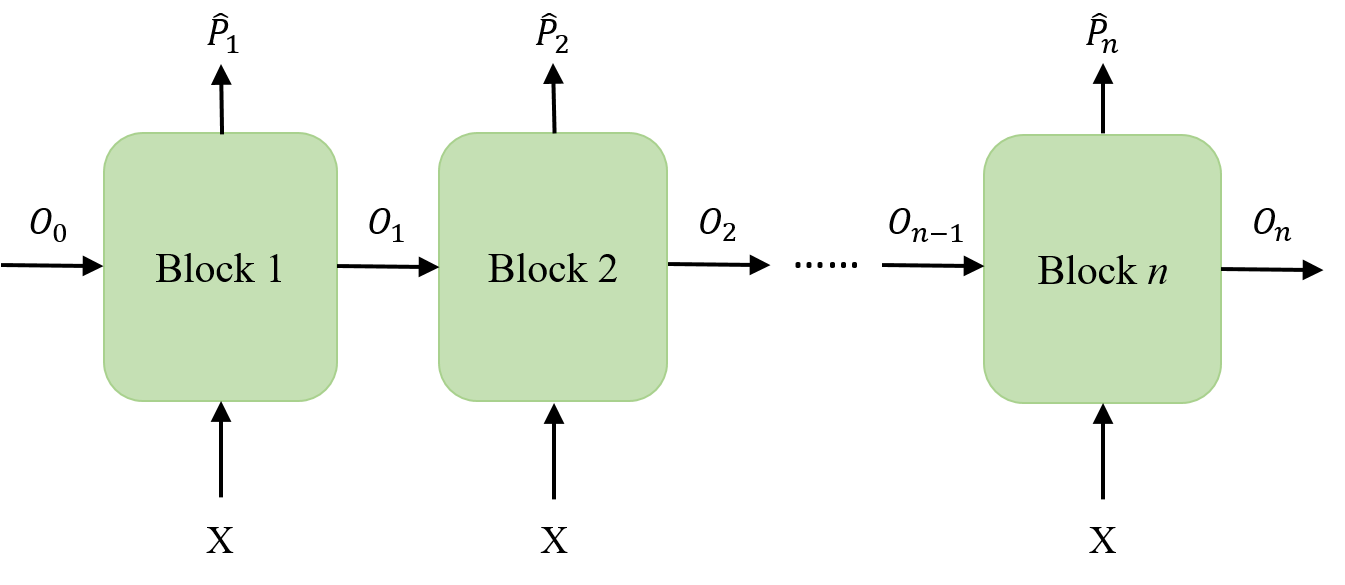}
\caption{The structure of LISA. It consists of $n$ blocks of a solution construction procedure (Block 1, $\cdots$, Block $n$).}\label{RNN}
\end{figure}

\subsection{Note}

We should emphasize that the framework used for building LISA is totally different from the unfolding structure of DetNet and its variants. In DetNet and its variants, the projected gradient descent is unfolded to several steps (layers). A neural network is designed to approximate the projection operator $\cal P$ in Eq.~(\ref{detnet}). The unfolding structure assembles the iterations of the projected gradient descent algorithm. Full solutions are updated along the unfolding.

In LISA, a series of neural networks are used to assemble a full solution. LISA concerns more about how a full solution to the detection problem is gradually constructed. Domain knowledge about constructing a solution can be readily incorporated in such a framework. On the other hand, the solution construction in each DetNet layer is limited to the design of neural networks.


\subsection{Training and Testing LISA}

In our experiments, the data used for training LISA is generated in two steps. First, we sample a set of channel matrix $\bm{H}$s. Each element of $\bm{H}$ is sampled from a normal distribution with zero mean and unit variance. For each $\bm{H}$, we generate data by applying Eq.~(\ref{problemx}). Note that the sampled $\bm{H}$ is not positive-definite ensured. The noise variance was randomly sampled so that the SNR will be uniformly distributed in $[\text{SNR}_{\min}, \text{SNR}_{\max}]$, where the SNR is the same as commonly defined in literature:
\begin{equation}
 \text{SNR}=\frac{\mathbb{E}||\bm{Hs}||_2^2}{\mathbb{E}||\bm{n}||_2^2}.
\end{equation}



Suppose the generated training data set is ${\cal D} =\{({\bm{s}}^{(i)},{\bm{H}}^{(i)},{\bm{y}}^{(i)}), i=1,2,\cdots,N\}$, where $N$ is the number of samples. ${\bm{s}}^{(i)}=\left(s_1^{(i)},s_2^{(i)},\cdots,s_{2N_T}^{(i)}\right)^\intercal$ is the $i$-th transmitted symbol vector, and ${\bm{y}}^{(i)}=\left(y_1^{(i)},y_2^{(i)},\cdots,y_{2N_T}^{(i)}\right)^\intercal$ is the observed symbol vector.

Before training, we perform QL-decomposition on the channel matrix of each sample
\begin{equation}{\bm{H}}^{(i)}={\bm{Q}}^{(i)}{\bm{L}}^{(i)}\end{equation}Multiplying $\left(\bm{Q}^{(i)}\right)^\intercal$ over $\bm{y}^{(i)}$, we obtain a converted data set $\widetilde{\cal D}=\{({\bm{s}}^{(i)},{\bm{L}}^{(i)},{\bm{\widetilde{y}}}^{(i)}), i=1,2,\cdots,N\}$, where ${\bm{\widetilde{y}}}^{(i)}={{\bm{Q}}^{(i)}}^\intercal {\bm{y}}^{(i)}$. $\widetilde{\cal D}$ will be used in the training of LISA.

The choice of the loss function is critical for neural network training. We regard the MIMO detection problem as a classification problem since all symbols take value from a discrete set. Therefore, the cross entropy loss is employed for training. Specifically, the loss function we used in the training process is written as:
\begin{equation}\label{loss}
 L(\bm{\theta}) = \frac{1}{N}\sum_{i=1}^{N}\sum_{j=1}^{2N_T}{\cal L}_{CE}\left(\bm{p}\left(s^{(i)}_j\right),\hat{\bm{p}}\left(s^{(i)}_j;\bm{\theta}\right)\right)
\end{equation}where ${\cal L}_{CE}(\cdot, \cdot)$ represents the cross entropy function:
\begin{equation}\small {\cal L}_{CE}\left(\bm{p}\left(s^{(i)}_j\right),\hat{\bm{p}}\left(s^{(i)}_j\right)\right) = -\sum_{k=1}^M \bm{p}_k\left(s^{(i)}_j\right) \log\hat{\bm{p}}_k\left(s^{(i)}_j;\theta\right)\label{lce} \end{equation}and $\bm{p}(s^{(i)}_j)$ represents the true probability distribution of the symbol $s^{(i)}_j$, $s^{(i)}_j$ denotes the $j$th symbol of the $i$th sample $\bm{s}^{(i)}$.
The probability distribution can be expressed as:
\begin{equation}
 \bm{p}(s^{(i)}_j)={
 \left[ \begin{array}{ccc}
 I(s^{(i)}_j=x_1)\\
 I(s^{(i)}_j=x_2)\\
 \cdots\\
 I(s^{(i)}_j=x_M)
 \end{array}
 \right ]}
\end{equation}where $I(\cdot)$ is the indicator function. Actually, $\bm{p}(s^{(i)}_j)$ is the one-hot encoding of the symbol $s^{(i)}_j$. $\hat{\bm{p}}(s^{(i)}_j;\bm{\theta})$ is the prediction of the probability distribution of the last block to $s^{(i)}_j$, and $\bm{\theta}$ is the parameter of LISA. The subscript $k$ in Eq.~(\ref{lce}) denotes the $k$th element of vector $\bm{p}(s^{(i)}_j)$ and $\hat{\bm{p}}(s^{(i)}_j;\bm{\theta})$, respectively. In the sequel, let
\[\hat{\bm{p}}^{(i)}(\bm{\theta}) =
\begin{bmatrix}
\hat{\bm{p}}_1(s^{(i)}_1;\bm{\theta}) & \cdots &\hat{ \bm{p}}_M(s^{(i)}_{1}; \bm{\theta}) \\
\vdots 					& \vdots & \vdots \\
 \hat{\bm{p}}_1(s^{(i)}_{2N_T};\bm{\theta}) & \cdots & \hat{ \bm{p}}_M(s^{(i)}_{2N_T}; \bm{\theta})
\end{bmatrix}
\] where each row represents the estimated signal distribution for the $j$th symbol ($j = 1, \cdots, 2N_T$) of the $i$th sample. Concisely, we write
\begin{equation}
\hat{\bm{p}}^{(i)}(\bm{\theta})= \text{LISA}(C_0,h_0,\bm{\widetilde{y}}^{(i)},\bm{L}^{(i)}; \bm{\theta}).
\end{equation}That is, LISA is considered as a parameterized function with appropriate input and output.

The pseudo code of training LISA is illustrated in Alg.~\ref{tlisa}, in which multiple epochs ($K$) are used. At each epoch, a number of $BN$ mini-batch of data are employed to optimize for the parameters $\bm{\theta}$ of LISA, which is randomly initialized to be $\bm{\theta}_0$ (line~\ref{t1}). For each mini-batch, first a set of $b$ data are randomly generated and transformed by QL-decomposition (lines~\ref{t2} to~\ref{t3}). Given the transformed data,  the estimated signals $\hat{\bm{p}}^{(i)}(\bm{\theta}_{t-1})$ for each problem $\bm{\widetilde{y}}^{(i)} = \bm{L}^{(i)}\bm{s}^{(i)}+\bm{\widetilde{n}}^{(i)}$ is obtained by applying LISA with current parameter $\bm{\theta}_{t-1}$ (line~\ref{t4}). The corresponding loss can be computed (line~\ref{t5}). The ADAM algorithm is then applied to update $\bm{\theta}_t$ (line~\ref{t6}). 

\begin{algorithm}[h]
 \caption{The Training Procedure for LISA.}\label{tlisa}
  \KwIn{$\text{the number of epoch}\ K$; $\text{the number of mini-batches used in each epoch}\ BN$; \text{the batch size}\ $b$; the parameter $\psi$ of Adam\;}
  \KwOut{$\text{the trained LISA}$}
 initialize the parameters of LISA $\bm{\theta}_0 \in \mathbb{R}^d, C_0\leftarrow \bm{0}$ and $h_0 \leftarrow \bm{0}$; and $t \leftarrow 0$\; \label{t1}
 \For {$k \leftarrow 1$ to $K$}
 {
    \For {$l \leftarrow 1$ to $BN$}
   {
       $t \leftarrow t+1$\;
       \Comment{data generation;}
       ${\cal D} \leftarrow \{({\bm{s}}^{(i)},{\bm{H}}^{(i)},{\bm{y}}^{(i)}), i=1,2,\cdots,b\}$\; \label{t2}
       \Comment{data transformation by QL decomposition;}
       ${\bm{H}}^{(i)}\leftarrow {\bm{Q}}^{(i)}{\bm{L}}^{(i)}$ and ${\bm{\widetilde{y}}}^{(i)}\leftarrow{{\bm{Q}}^{(i)}}^\intercal {\bm{y}}^{(i)}$ for $i = 1,\cdots,b$\;
       obtain $\widetilde{\cal D}\leftarrow \{({\bm{s}}^{(i)},{\bm{L}}^{(i)},{\bm{\widetilde{y}}}^{(i)}), i=1,2,\cdots,b\}$\; \label{t3}
       \Comment{forward propagation;}
       $\hat{\bm{p}}^{(i)}(\bm{\theta_{t-1}})= \text{LISA}(C_0,h_0,\bm{\widetilde{y}}^{(i)},\bm{L}^{(i)}; \bm{\theta}_{t-1})$ for $i=1,2,\cdots,b$\; \label{t4}
       calculate $ L(\bm{\theta}_{t-1})$ (cf. Eq.~(\ref{loss}))\; \label{t5}
       \Comment{back propagation;}
      $\bm{\theta}_t \leftarrow \text{ADAM}(L(\bm{\theta}_{t-1}); \psi)$; \label{t6}
   }

 \Return the trained LISA with parameter $\bm{\theta}_t$}
\end{algorithm}

In the testing phase, suppose that the new observed signal is $\bm{y}=(y_1,y_2,\cdots,y_{2N_R})^\intercal$ and its corresponding channel matrix is $\bm{H}=[ \bm{H}_1,\bm{H}_2,\cdots,\bm{H}_{2N_T}]$, where $\bm{H}_i$ denotes the $i$-th column of $\bm{H}$. To recover the transmitted signal, we first perform QL-decomposition on $\bm{H}$ to convert the received signal to $\widetilde{\bm{y}}=\bm{Q}^\intercal\bm{y}$. We then feed the trained LISA with $\widetilde{\bm{y}},\bm{L}$. The transmitted signal is recovered as
\begin{equation}\hat{s}_i  =\arg\max\,\hat{\bm{p}}_i, 1\leq i \leq 2N_T\end{equation}where $\hat{\bm{p}}_i $ is the output of the $i$-th component of LISA.

\subsection{Complexity and Memory Cost Analysis}

In this section, the computational complexity and memory cost of LISA are analysed.

The computational complexity of LISA consists of two parts, the QL decomposition and the computation of latent variables $C_{{k}}, h_{{k}}, 1\leq {{k}} \leq 2N_T$ in LSTM. The complexity of the QL decomposition is $\mathcal{O}(N_RN_T^2)$. The updating $C_{{k}}$ and $h_{{k}}$ (cf. Eq. (12)) involves calculating intermediate variables $f_{{k}}, i_{{k}}, \tilde{C}_{{k}}$ and $o_{{k}}$. The computation of each has a complexity $\mathcal{O}((d_h + d_i^{{k}})d_h)$, where $d_h$ and $d_i^{{k}}$ represents the dimension of $C_{{k}}$ and $x_{{k}}$, respectively. Note $d_i^{{k}}$ differs at each step ${{k}}$, its value is ${{k}}+1$ for ${{k}} = 1, \cdots, 2N_T$. In total, the computational complexity of LISA is $\mathcal{O}(N_T^2N_R + N_Td_h^2 + N_T^2d_h)$.

In comparison, the expected computational complexity of SD is known to be $\mathcal{O}(M^{\beta N_T})$~\cite{1408197,1468474}, where $\beta \in (0,1]$ is a factor depending on the noise level, while the complexity of MMSE is $\mathcal{O}(N_T^3 + N_T^2N_R)$. It is obvious that the complexity of LISA is significantly smaller than SD and MMSE in case of large $N_T$. {On the other hand, the complexity of learning-based detectors, including MMNet and OAMP-Net, is $O(N_RN_T^2)$ and  $O(N_R^3)$, respectively.  It is seen that LISA has a higher complexity than these detectors. The complexity of DetNet is $O(N_RN_T^2+d_ZN_T+d_Zd_V)$ where $d_Z$ and $d_V$ are some dimensions of introduced augmented variables. The comparison of complexity between LISA and DetNet depends on the settings of $d_h$ and $d_Z$ and $d_V$.

We further show the number of parameters in LISA. In LISA, at time step $t$, the number of parameters in LSTM is $4(d_hd_i^t + d_h^2 +d_h)$  and $d_h\cdot M$ for the Softmax layer where $M$ is the dimension of the output $\hat{\bm{p}}_t$ (i.e. the number of constellation). Therefore, the total number of parameters in LISA is $4N_Td_h(2N_T + 2d_h + 5) + 2N_Td_hM$. Regarding the memory cost of LISA, it is seen that only the hidden states $C_t$ and $h_t$ in LSTM need to be stored during training and testing. Therefore, the memory cost of LISA at each time step $t$ is $2d_h$. Furthermore, if a real number requires $B$ bits to store, then the total memory cost of LISA is $2d_hB$ bits.

\section{Experimental Results}\label{experiments}

In this section, comparison results of LISA against existing detectors are provided in both the fixed channel and varying channel scenarios. The code is implemented in Pytorch. The computing platform is two 14-core 2.40GHz Intel Xeon E5-2680 CPUs (28 Logical Processors) under Windows 10 with GPU NVIDIA Quadro P2000. The ADAM optimizer~\cite{kingma2014adam} is used to train LISA with the following hyper-parameter settings: ${\beta}_1 = 0.9, {\beta}_2 = 0.999, lr = 0.0006$ and $\epsilon = 10^{-8}$. Please see Appendix for details of ADAM.

The compared detectors include two linear detectors (ZF and MMSE), one machine learning based algorithm (two-stage ADMM), two learning to learn methods (DetNet~\cite{8227772} and MMNet~\cite{Shirkoohi2019AdaptiveNS}), the sphere decoder (SD) and the MLD. Specifically, the hyper-parameters used in DetNet and MMNet are the same as those mentioned in the original papers, respectively. The SD presented in~\cite{5773010} is used for comparison. The bit error rate (BER) is used to measure the performance of these detectors.

\subsection{Results in the fixed channel scenario}

In the fixed channel scenario, the same channel matrix $\bm{H}$ is applied in training and testing. The structure of the DNN is a fully connected neural network with three layers, and the number of neurons in the hidden layer is 600.

In the training, for a fixed $\bm{H}$, ten million samples are generated and used to train LISA. The training samples consist of signals of different SNR values in $[2, 8]$, which are sampled uniformly. Following standard deep neural network training procedure, we train LISA in 10 epochs. At each epoch, the mini-batch training method is adopted with batch size 1000.  The BER of the learned LISA on a test data set with size 100K, which is generated the same as the training data, is compared among the detectors.

The comparison result regarding the fixed channel with QPSK modulation is shown in Fig.~\ref{fixed_QPSK}. It can be seen that LISA performs much better than the linear detectors (MMSE, ZF) and even achieve near-MLD performance. The {fully-connected (FC) network used in}~\cite{DBLP:journals/corr/abs-1805-07631} has very poor performance: it performs even worse than the linear detectors, while MMNet performs comparably to MMSE. The performance of the two-stage ADMM is slightly better than {the FC network}, but still worse than MMSE and ZF at high SNR levels. {Although DetNet performs better than MMSE, there is still a big gap from MLD}. The SD achieves near optimal performance, but its time complexity is exponential to low SNR levels and the number of transmitting antennas~\cite{1408197,1468474}.}
\begin{figure}[htbp]
\centering
\includegraphics[scale=0.5]{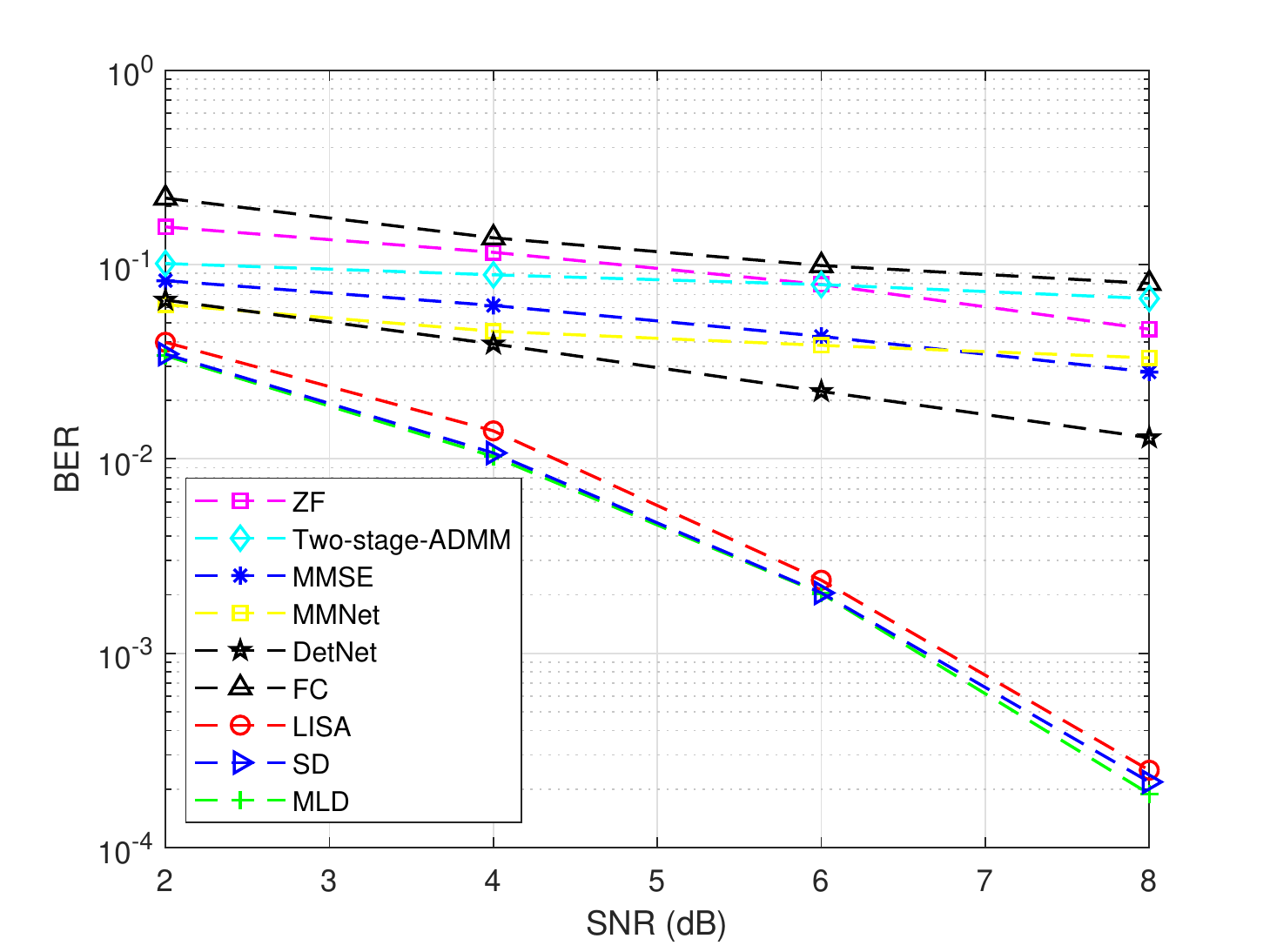}
\caption{Comparison of the detection algorithms' BER performance in the $4 \times 4$ fixed channel case over QPSK modulated signals.}
\label{fixed_QPSK}
\end{figure}

\subsection{Results in the varying channel scenario}


In this experiment, 100 million training samples are used, in which a range of SNR values are used to generate these samples. When generating the samples, we do not exclude $\bm{H}$s that are ill-conditioned (with large condition number) or even singular. The dimension of the hidden state, i.e. $C_t$, is set as 600. Two blocks are used in LISA. In the training phase, we train LISA with 40 epochs and the batch size in each epoch is 20K.


We found the following post-processing procedure is able to further improve the performance of LISA. That is, for each $\bm{H}$ and an observation $\bm{y}$, signals are recovered in original order and reverse order in the testing phase. The transmitted signal is then determined based on the two predictions.

In the original order, the signal is recovered from $1$ to $2N_T$. The lower-triangle matrix obtained by QL decomposition on $\bm{H}$ and $\bm{y}$ is input to the learned LISA. In the reverse order, the signal is recovered from $2N_T$ to $1$. To do so, we first reverse the order of the columns of the channel matrix to obtain $\bm{H}'=[\bm{H}_{2N_T},\cdots,\bm{H}_2,\bm{H}_1]$. The QL-decomposition on $\bm{H}'$ ($\bm{H}'=\bm{Q}'\bm{L}'$) is then used as the input to LISA. 

Suppose the recovered signal in the original order is $\bm{s'}$ and that of the reversed order is $\bm{s}''$. Combing the two recovered signals, the transmitted signals are recovered as \begin{equation}\bm{s}^* = \arg\min\limits_{{\bm{s}}\in\{{\bm{s}',\bm{s}''}\}}||\bm{y}-\bm{H}\bm{s}||_2^2.\end{equation}That is, the signal with smaller recovery error is considered as the final recovered signal.
\begin{figure}[htbp]
\centering
\includegraphics[scale=0.5]{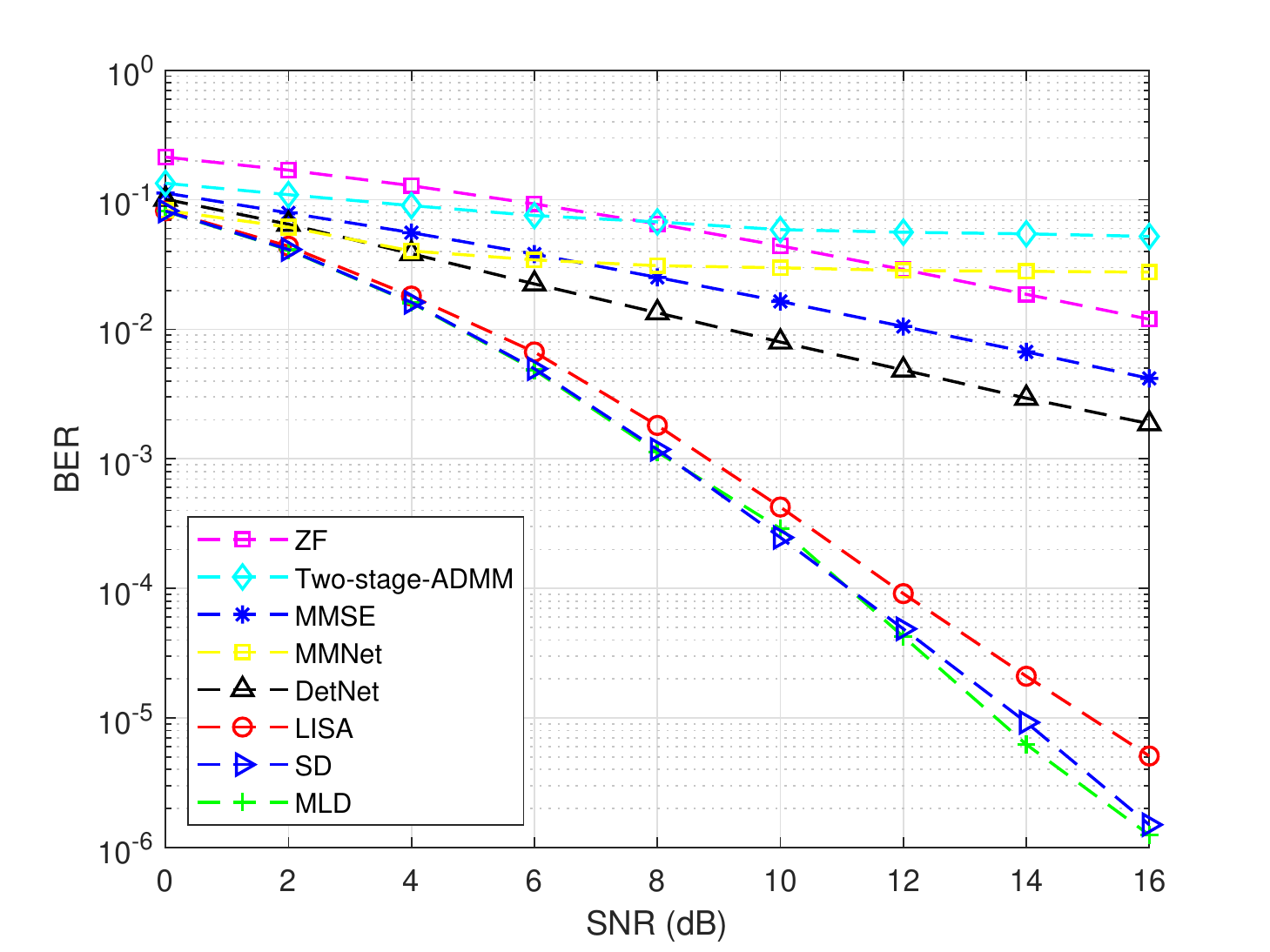}
\caption{Comparison of the detection algorithms' BER performance in the $4 \times 4$ varying channel case over QPSK modulated signal.}
\label{varying_QPSK}
\end{figure}

Fig.~\ref{varying_QPSK} shows the BER performance of the compared algorithms in varying channel with the QPSK modulation, where the complex channel matrix size is $4 \times 4$, i.e. $N_T = 4$ and $N_R = 4$. From Fig.~\ref{varying_QPSK}, it is seen that LISA performs extremely well in this case. Its BER performance is much lower than the compared algorithms, even close to the MLD. To the best of our knowledge, there is no such detectors that are able to reach the MLD performance. For the compared detectors, DetNet is only able to perform better than MMSE, while the two-stage ADMM performs worse than MMSE, a bit better than ZF. MMNet performs better than MMSE at low SNR levels. However, the performance of MMNet degrades severely and even worse than ZF at high SNRs.
\begin{figure}[htbp]
\small
\centering
\includegraphics[scale=0.43]{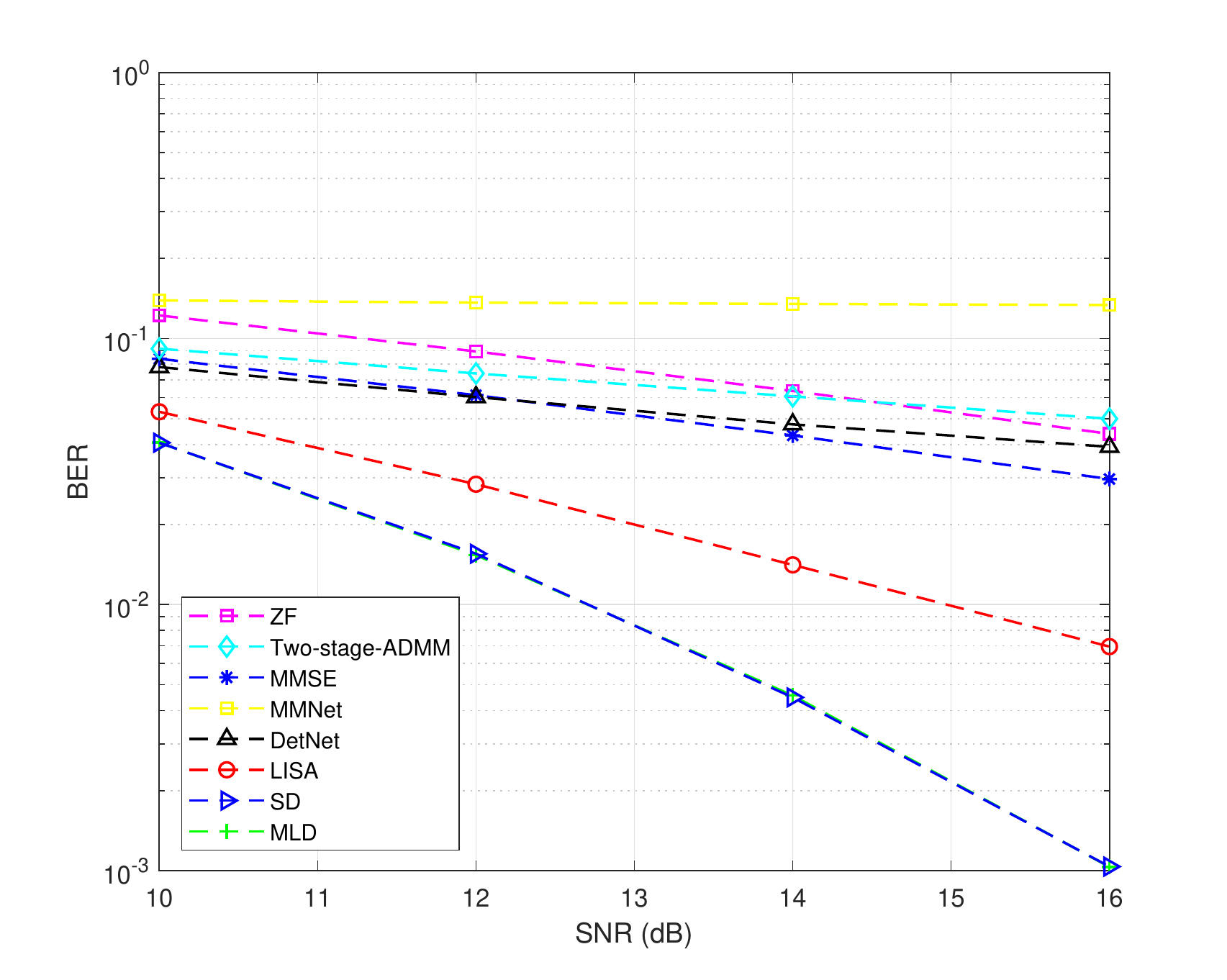}
\caption{Comparison of the detection algorithms' BER performance in the $4 \times 4$ varying channel case over 16-QAM modulated signal.}
\label{16QAM}
\end{figure}

Fig.~\ref{16QAM} shows the performance of LISA in a $4 \times 4$ MIMO system with 16-QAM modulation in terms of BER. It is seen that when the modulation order becomes higher, the detection performance of DetNet becomes similar to MMSE, but LISA still achieves better performance than MMSE. DetNet performs only comparably with MMSE, while two-stage ADMM performs only better than ZF, but worse than MMSE. It is seen that MMNet performs the worst in this case, which indicates that the performance of MMNet heavily depends on the modulation.
\begin{figure}[htbp]
\small
\centering
\includegraphics[scale=0.46]{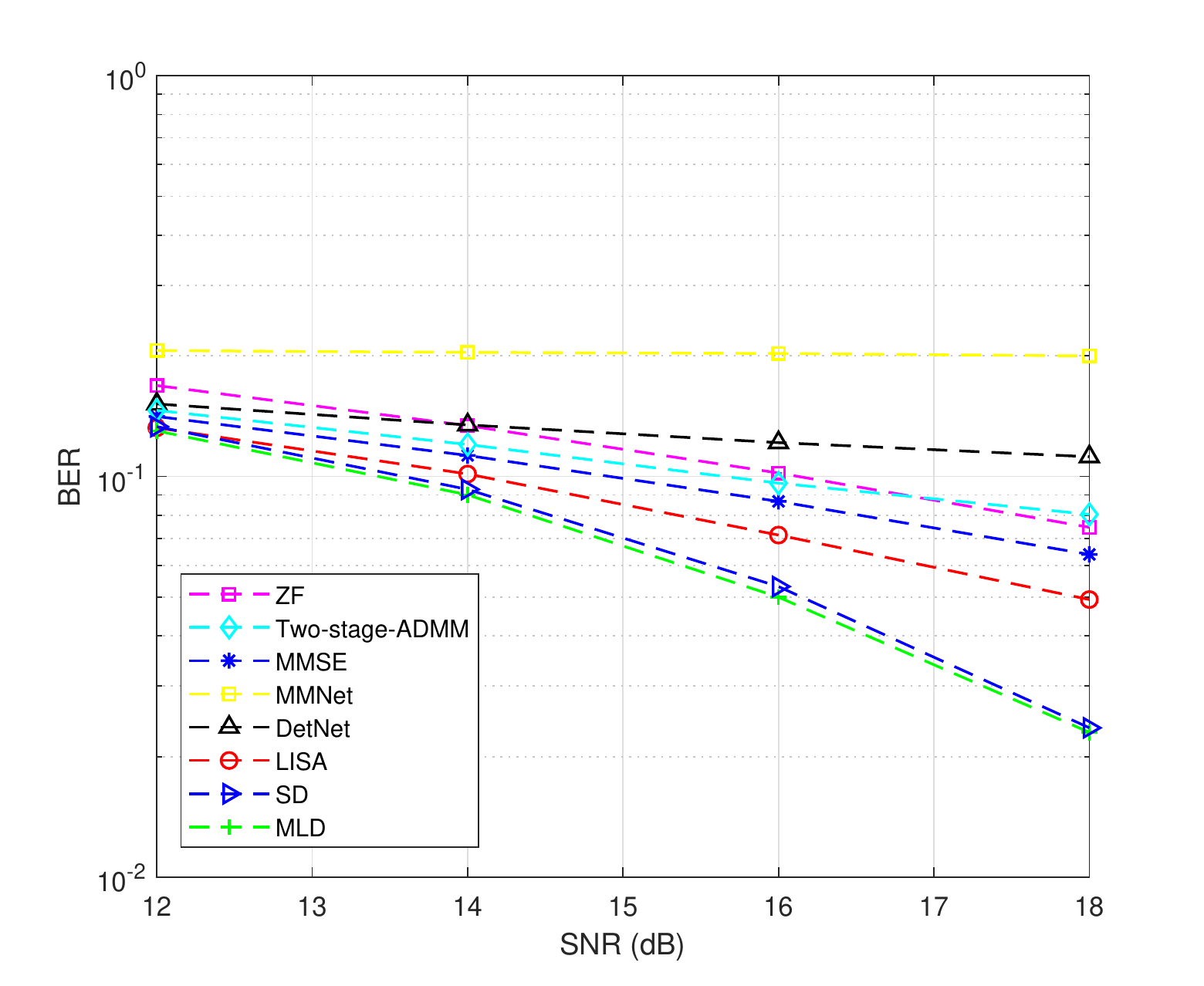}
\caption{Comparison of the detection algorithms' BER performance in the $4 \times 4$ varying channel case over 64-QAM modulated signal.}
\label{64QAM}
\end{figure}

The BER performance of LISA with 64-QAM modulation in $4 \times 4$ varying channel is shown in Fig.~\ref{64QAM}. From the figure, it is clear that MMNet performs the worst. DetNet performs better than MMNet. Actually, DetNet's performance is exactly the same as ZF. Two-stage ADMM performs a little worse than MMSE. LISA still performs better than MMSE, and it performs clearly the best among the compared detectors.

\begin{figure*}[htbp]
\small
\centering
\includegraphics[scale=0.4]{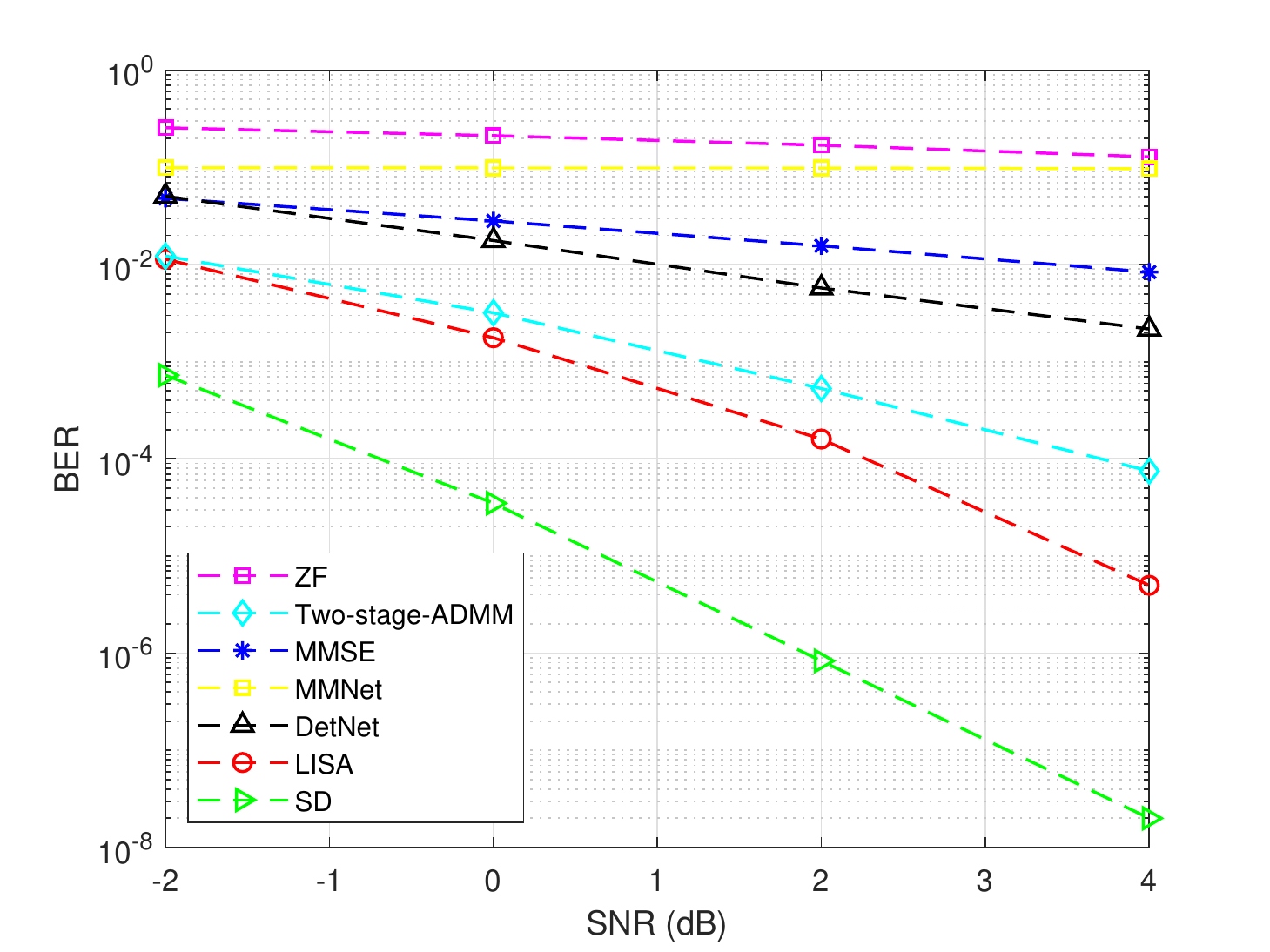}
\includegraphics[scale=0.4]{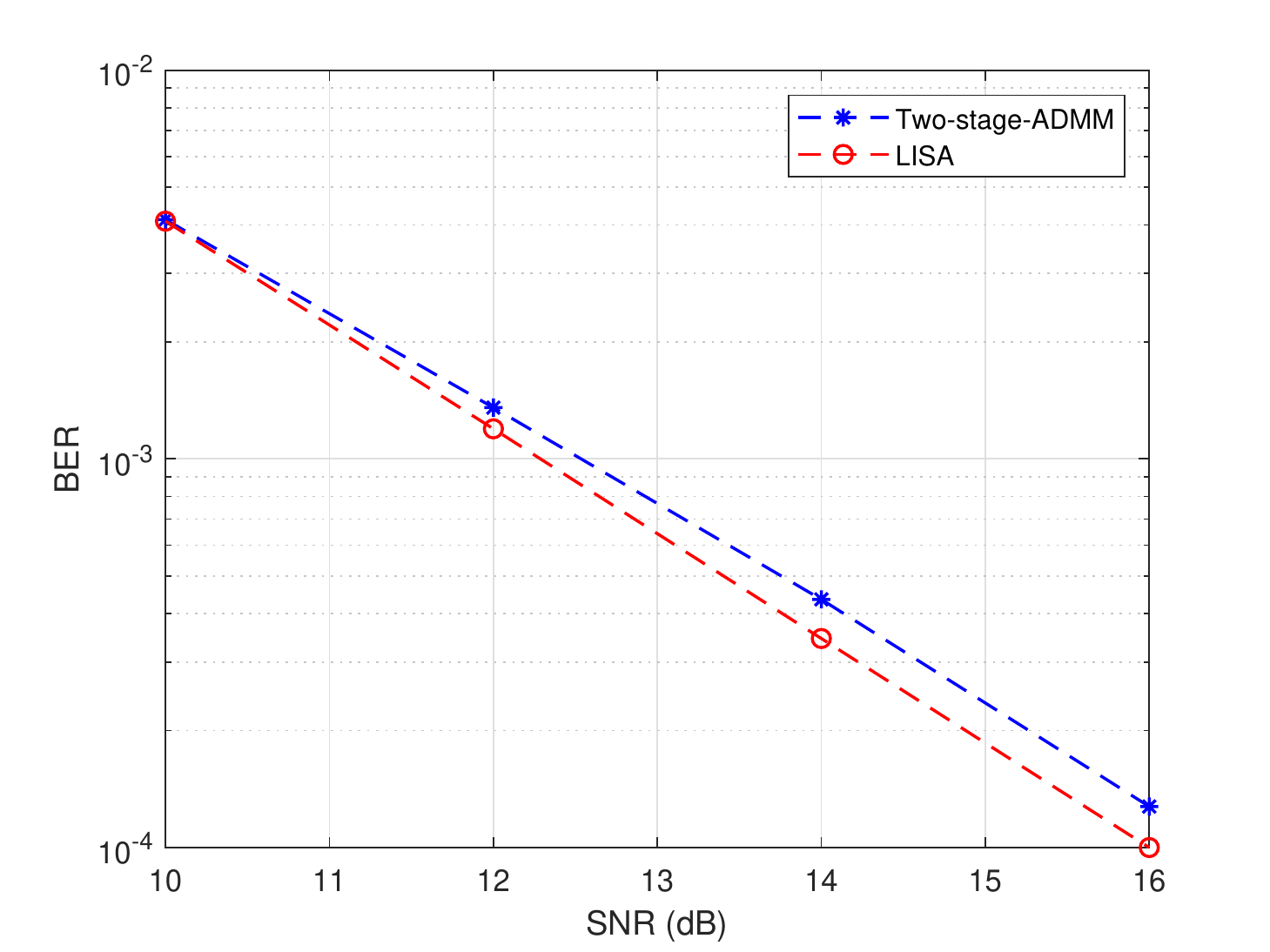}
\caption{Comparison of the detection algorithms' BER performance in $20 \times 20$ varying channel case over QPSK and 16-QAM modulated signals.}
\label{massive}
\end{figure*}

LISA can also be used for massive MIMO detection. Fig.~\ref{massive} shows the results of the compared detectors in $20\times20$ varying channel with QPSK and 16-QAM modulation. It is seen from the figure that LISA performs the best among the detectors, two-stage ADMM is worse, but better than DetNet in QPSK modulation. The performance of MMSE and ZF are very poor. MMNet performs better than ZF but worse than MMSE. With 16-QAM modulation, the performance of DetNet, MMNet, MMSE and ZF is no comparable to LISA and two-stage-ADMM, hence they are not shown in the figure.

\subsection{Further Results on Correlated Channel Matrices}

So far all the results are obtained by training LISA on channel matrices with independently sampled normal entities. To further test the performance of LISA against known detectors, in this section, LISA is trained on channel matrices whose entities are with correlations. The Kroneker model proposed in~\cite{1683382} is applied to generate the correlated channel matrices. In the Kronecker model, the channel matrix $\bm{H}$ is obtained as follows:
\begin{equation}\bm{H} = \bm{R}_r^{\frac{1}{2}}\bm{H}_w \bm{R}_t^{\frac{1}{2}}\end{equation}where $\bm{H}_w$ is an i.i.d. Rayleigh fading matrix, $\bm{R}_t$ ($\bm{R}_r$) is the covariance matrix in the transmitting (receiving) end:
\begin{equation}\bm{R}_r = \mathbf{I}, \qquad \bm{R}_t =
\begin{pmatrix}
1 		& \alpha 	& \cdots 	& 	\alpha \\
\alpha	& 1		& \cdots	& 	\alpha \\
\vdots  	& \vdots  	& \ddots 	& 	\vdots  \\
\alpha	& \cdots	& \alpha	&	1
\end{pmatrix}
\end{equation}where $\alpha$ is the correlation coefficient.


Fig.~\ref{correlated} shows the BER performance of LISA against the compared detectors in case of different correlation coefficients. From the figure, it is clearly seen that LISA performs significantly better than the rest of the algorithms. Especially, for the QPSK modulation, LISA obtains near-MLD performance. DetNet performs generally similar to MMSE, while two-stage-ADMM performs even worse than ZF in case of 16-QAM modulation. Two-stage-ADMM {and MMNet} is only slightly better than ZF in QPSK.

\begin{figure*}[htbp]
\small
\centering
\includegraphics[scale=0.375]{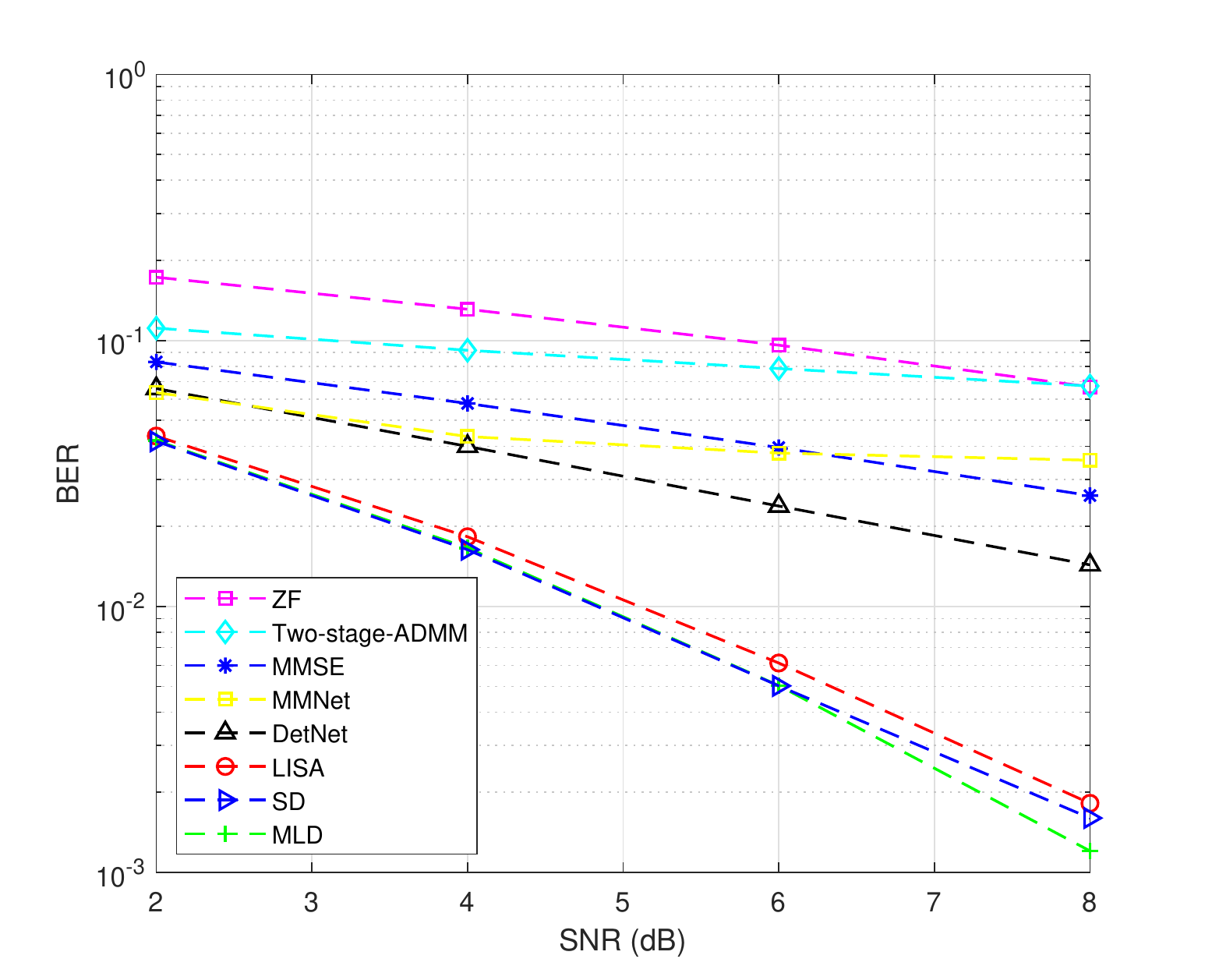}\includegraphics[scale=0.37]{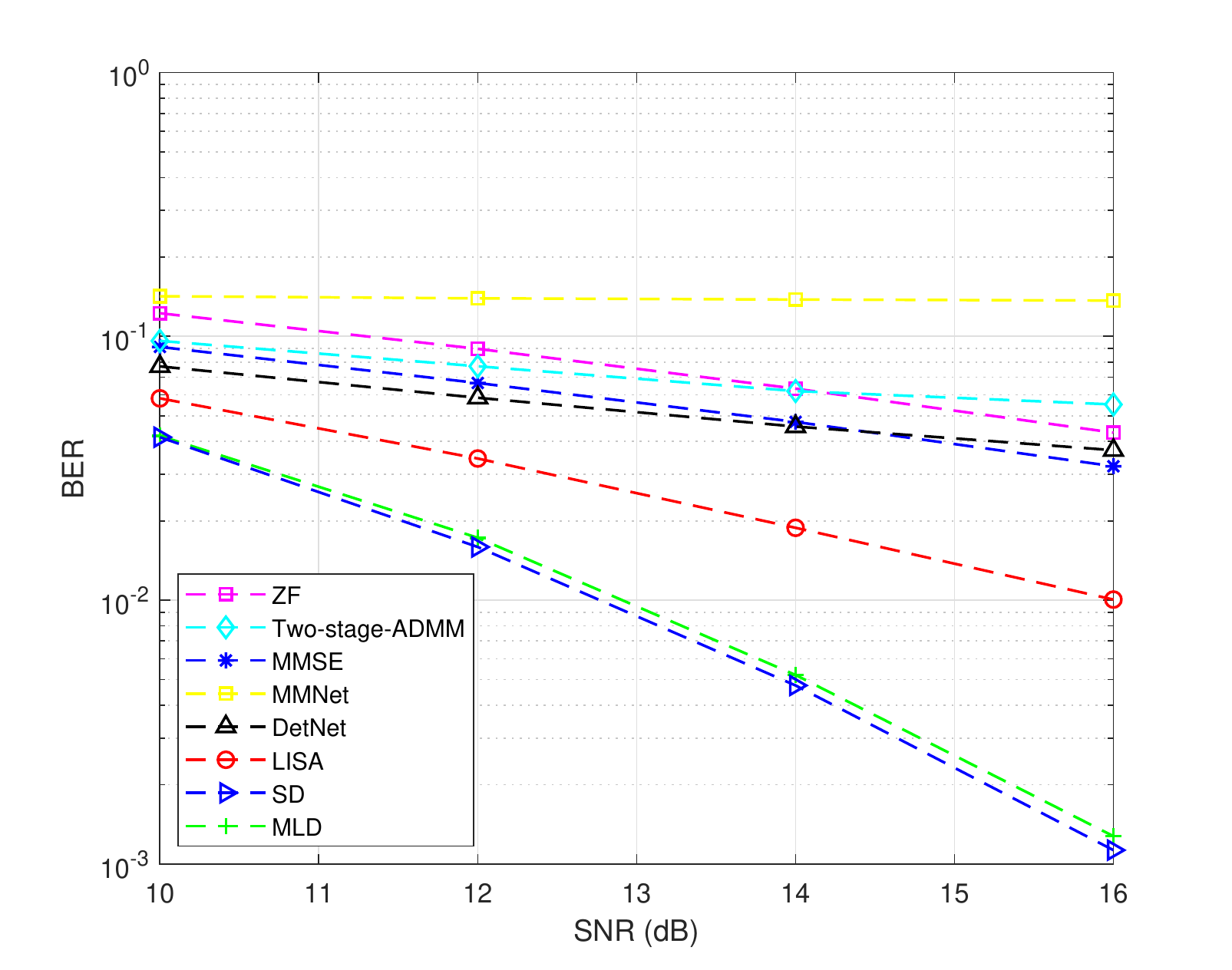}
\includegraphics[scale=0.4]{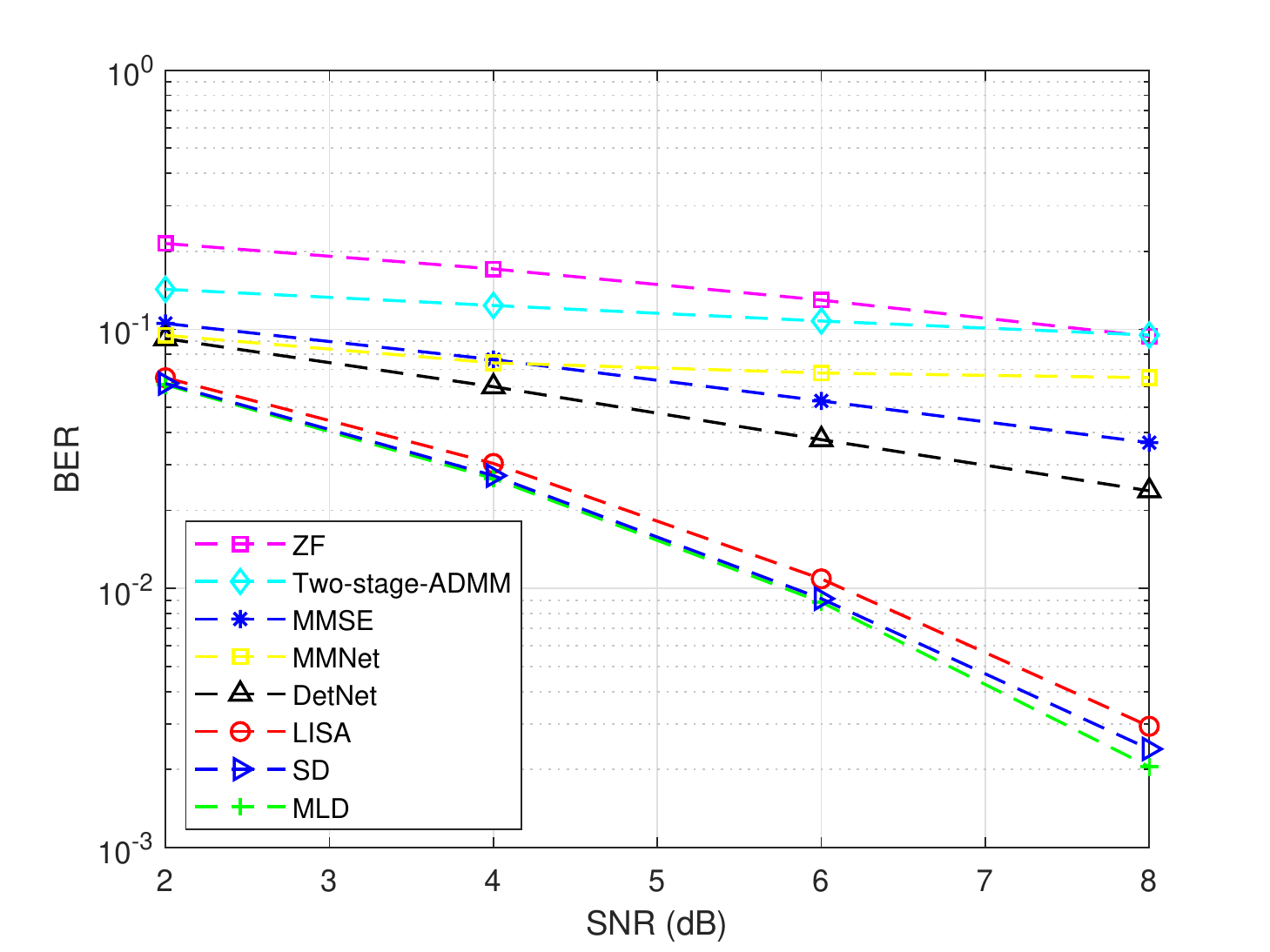}\includegraphics[scale=0.4]{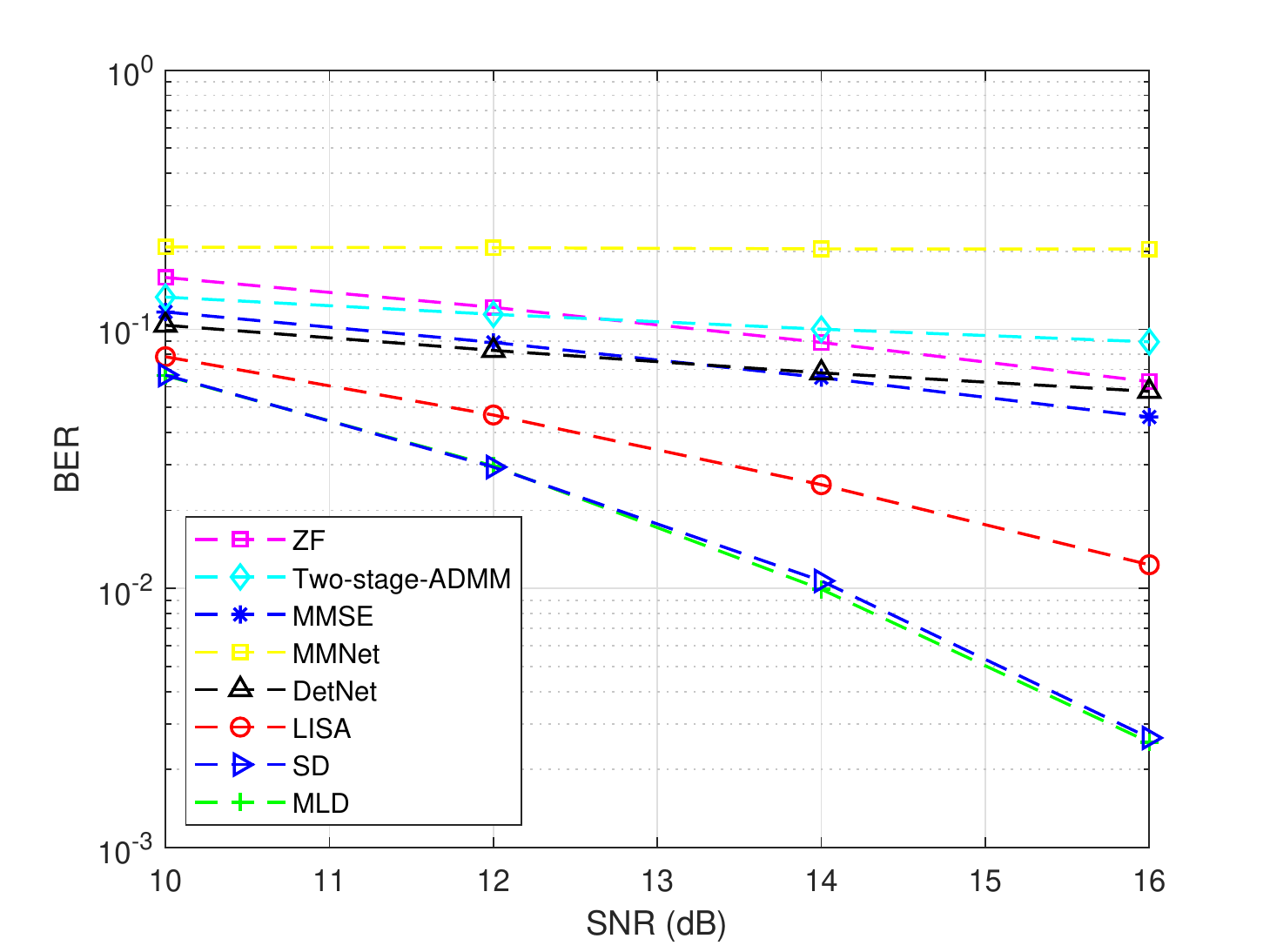}
\includegraphics[scale=0.4]{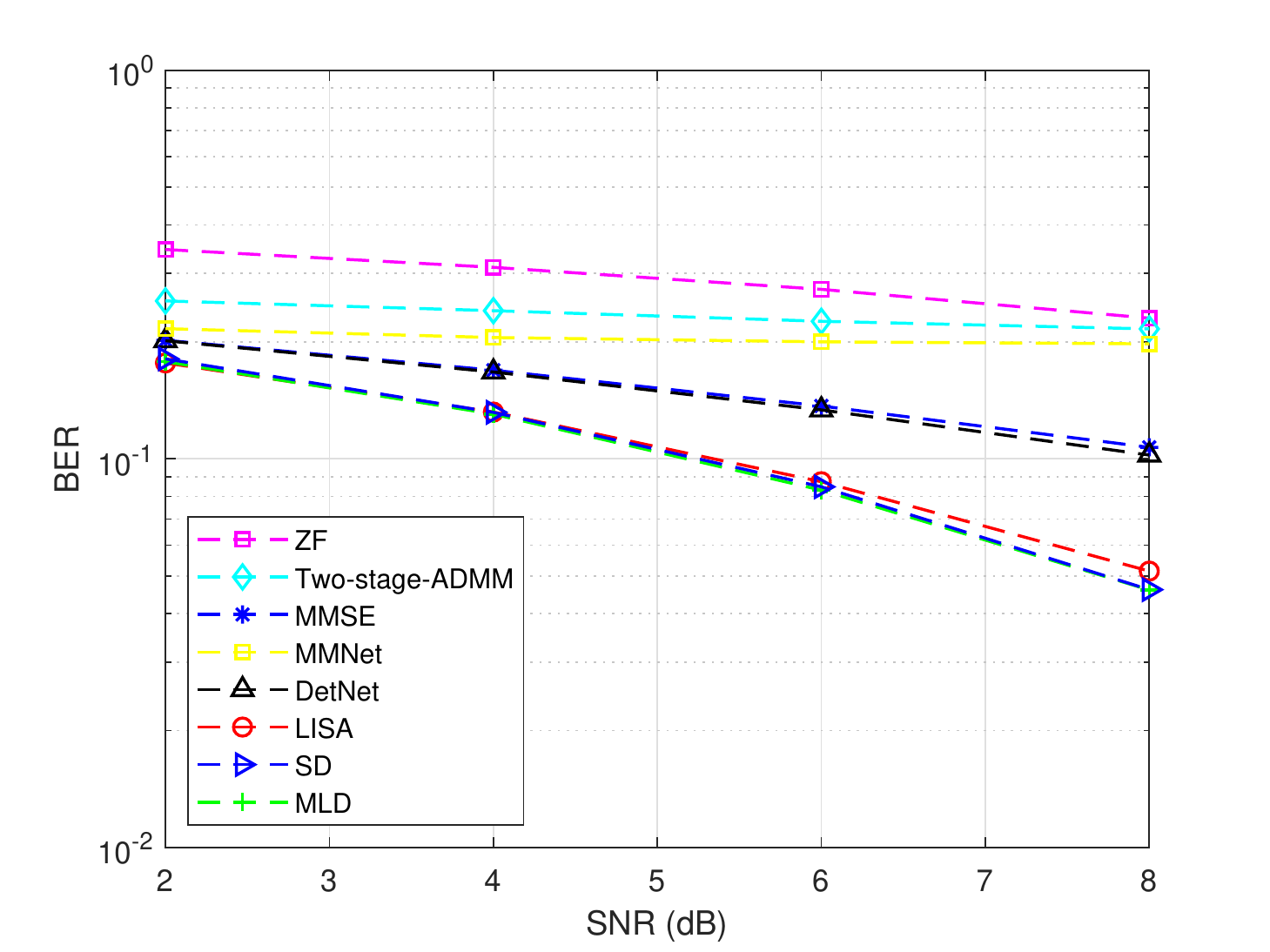}\includegraphics[scale=0.4]{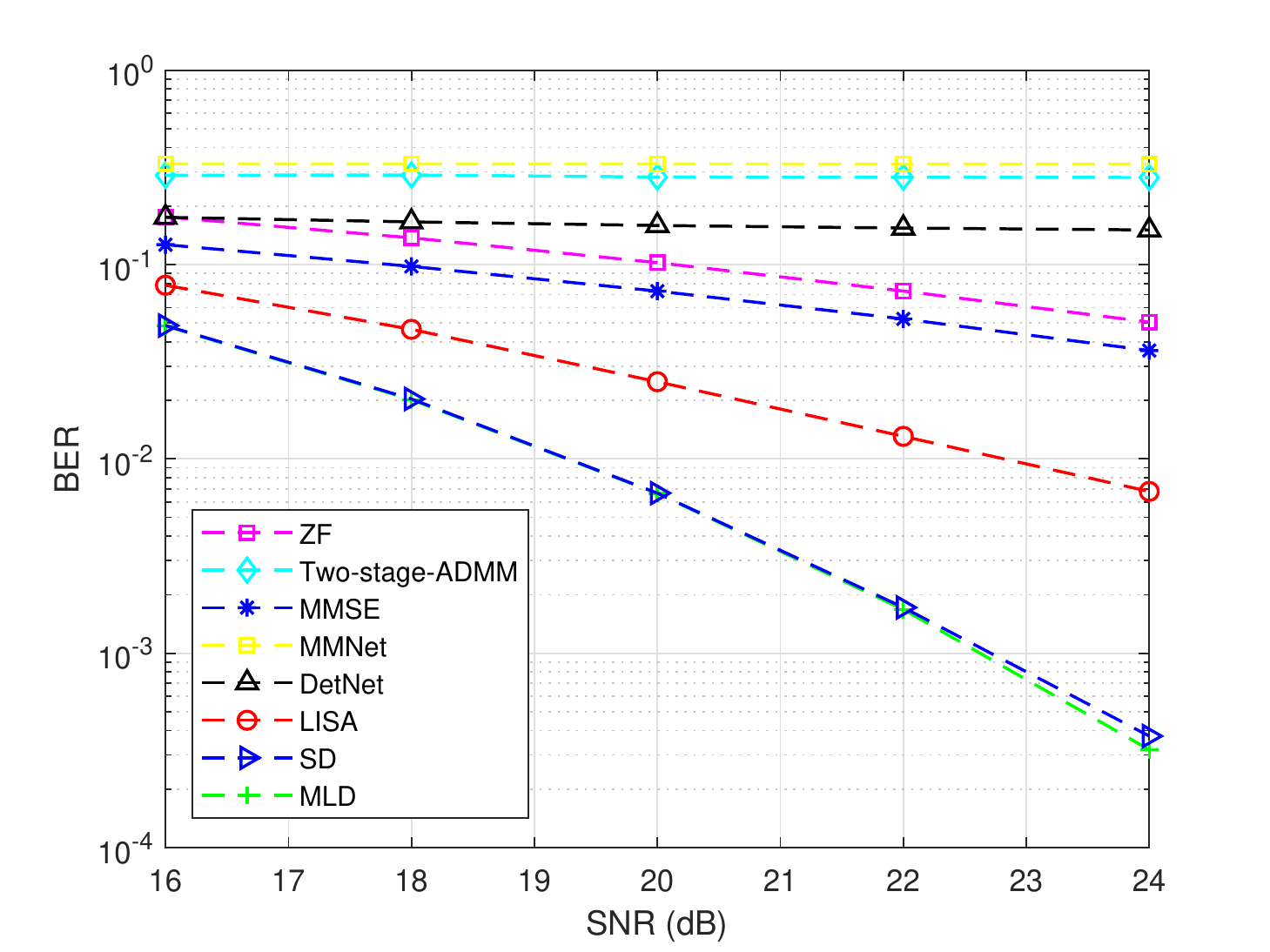}
\caption{The BER performance for $4 \times 4$ varying channel scenario with QPSK and 16-QAM modulation for channel matrices with different correlations. The correlation coefficients are 0.1, 0.5 and 0.9, respectively, from top to bottom.}
\label{correlated}
\end{figure*}

In summary, we may conclude that LISA works significantly better than all classical detectors (such as ZF, MMSE) and machine learning based algorithms (DetNet and two-stage-ADMM). Further, in comparison between the framework used by DetNet and LISA, we may conclude that the LISA framework can perform better than unfolding based learning to learn approach.

\subsection{Results on Channels with Imperfect CSI}

Notice that the above results are all obtained on channels with perfect CSI. In this section, we investigate LISA on channels with imperfect CSI.

In our implementation, the estimated channel matrix $\widehat{\bm H}$, assumed to be \begin{equation}\widehat{\bm H} = {\bm H} + \Delta {\bm H}\label{imper}\end{equation}where ${\bm H}$ is the actual channel matrix with i.i.d Rayleigh fading and $\Delta {\bm H}\sim \mathcal{CN}(0,\sigma_e^2I)$ is the CSI error~\cite{8398483}, is used in both the training and testing stages. $\sigma_e^2=0.1$ in the simulation experiment.


To simulate the imperfect CSI scenario, in the training stage, the observed signal $\bm y$ is first obtained using the actual channel matrix $\bm H$. But $\bm y$ is used to train LISA together with an imperfect CSI $\widehat{\bm H}$ sampled according to Eq.~\ref{imper}. In the testing stage, the data are sampled as the same way, and $\widehat{\bm H}$ is used to test the trained LISA.



The normality of $\Delta{\bm H}$ can be justified as follows. Let ${\bm S}_p$ denote the pilot signal matrix with size $N_T \times L_p$, where $L_p$ is the number of pilot signals and ${\bm R}_p$ is the received signal matrix with size ${N}_R \times L_p$. Then we have ${\bm R}_p = {\bm H}{\bm S}_p +{\bm N}_p$, where ${\bm N}_p$ is the Gaussian white noise. Using least square, $\bm H$ can be estimated as $\widehat{\bm H} = ({\bm R}_p{\bm S}_p^\mathrm{T})({\bm S}_p{\bm S}_p^\mathrm{T})^{-1}$. Therefore, the estimation error is $\Delta {\bm H} = \widehat{\bm H}-{\bm H} = {\bm N}_p{\bm S}_p^\mathrm{T}({\bm S}_p{\bm S}_p^\mathrm{T})^{-1}$. Since ${\bm N}_p$ follows the Gaussian distribution, so does $\Delta {\bm H}$.


Fig.~\ref{imperfect CSI} shows the BER performance of different detectors on channels with estimated CSI over QPSK modulation. From the figure, it is seen that LISA still achieves near-MLD performance. We thus may conclude that LISA is robust to imperfect CSI.  
\begin{figure}[htbp]
\small
\centering
\includegraphics[scale=0.43]{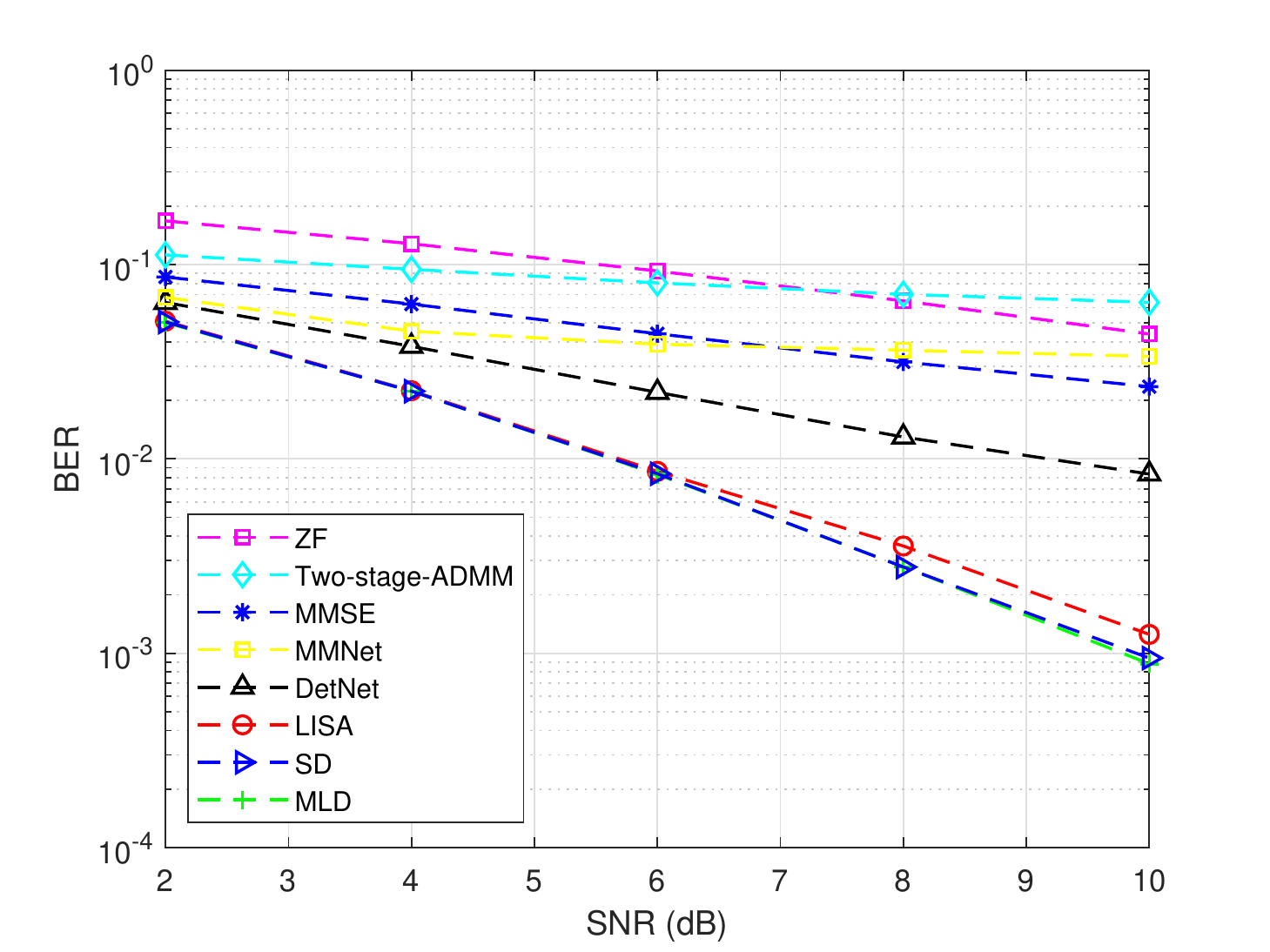}
\caption{Comparison results of LISA's BER performance in the $4 \times 4$ varying channel with imperfect CSI over QPSK modulation.}
\label{imperfect CSI}
\end{figure}

\subsection{Generalisation Analysis}

In this section, we analyse LISA's generalisation ability against correlation in the channel matrix and SNR. The experiment is designed similarly to that in~\cite{DBLP:journals/corr/abs-1907-09439}.

In all the experiments carried out above, the training data and test data are sampled from the same scenario. However, it is interesting to know whether LISA is able to perform well in the scenarios that have not been trained for.

To answer this question, we train LISA with the training data drawn from correlated channel with correlation coefficient $\alpha = 0.5$ and $\text{SNR} = 2$dB, but test LISA on correlated channels with $ \alpha = 0.1, 0.5, 0.9$ and SNR from 2dB to 8dB. With such mismatched training and test data, we are able to analyse the impact of mismatched data on LISA's performance. The experimental results in QPSK modulation are shown in Fig.~\ref{robustness}, in which LISA's BER values with matched data and with mismatched data are given against SNRs.
\begin{figure}[htbp]
\small
\centering
\includegraphics[scale=0.3]{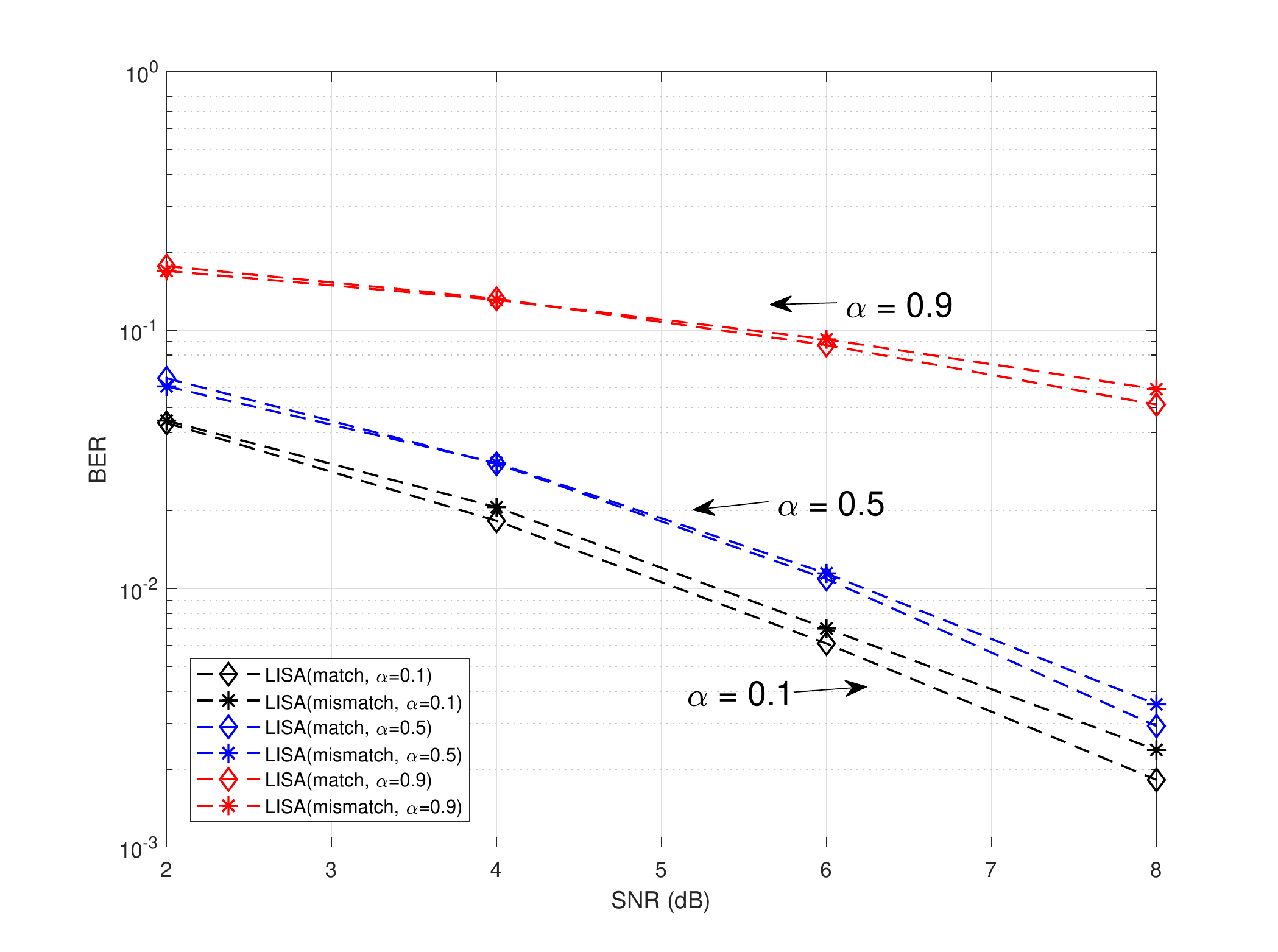}
\caption{Comparison results of LISA's BER performance in the $4 \times 4$ varying channel with correlation and SNR mismatches over QPSK modulation.}
\label{robustness}
\end{figure}

From Fig.~\ref{robustness}, it is seen that LISA's performance degrades a little bit when the SNR and correlation coefficient of training data and test are mismatched, but the gap is very small. We may conclude that LISA is robust against correlation coefficient and SNR.

\subsection{Sensitivity Analysis}

In LISA, one of the important architecture parameters is the number of neurons used in LSTM. We investigate how the number affects the performance of LISA in the varying channel scenario. Fig.~\ref{nodes} shows the performance of LISA in terms of different number of neurons. From the figure, it is seen that along the increasing of neurons, the performance of LISA becomes better. However, the performance improvement becomes less significant.
\begin{figure}[htbp]
\small
\centering
\includegraphics[scale=0.4]{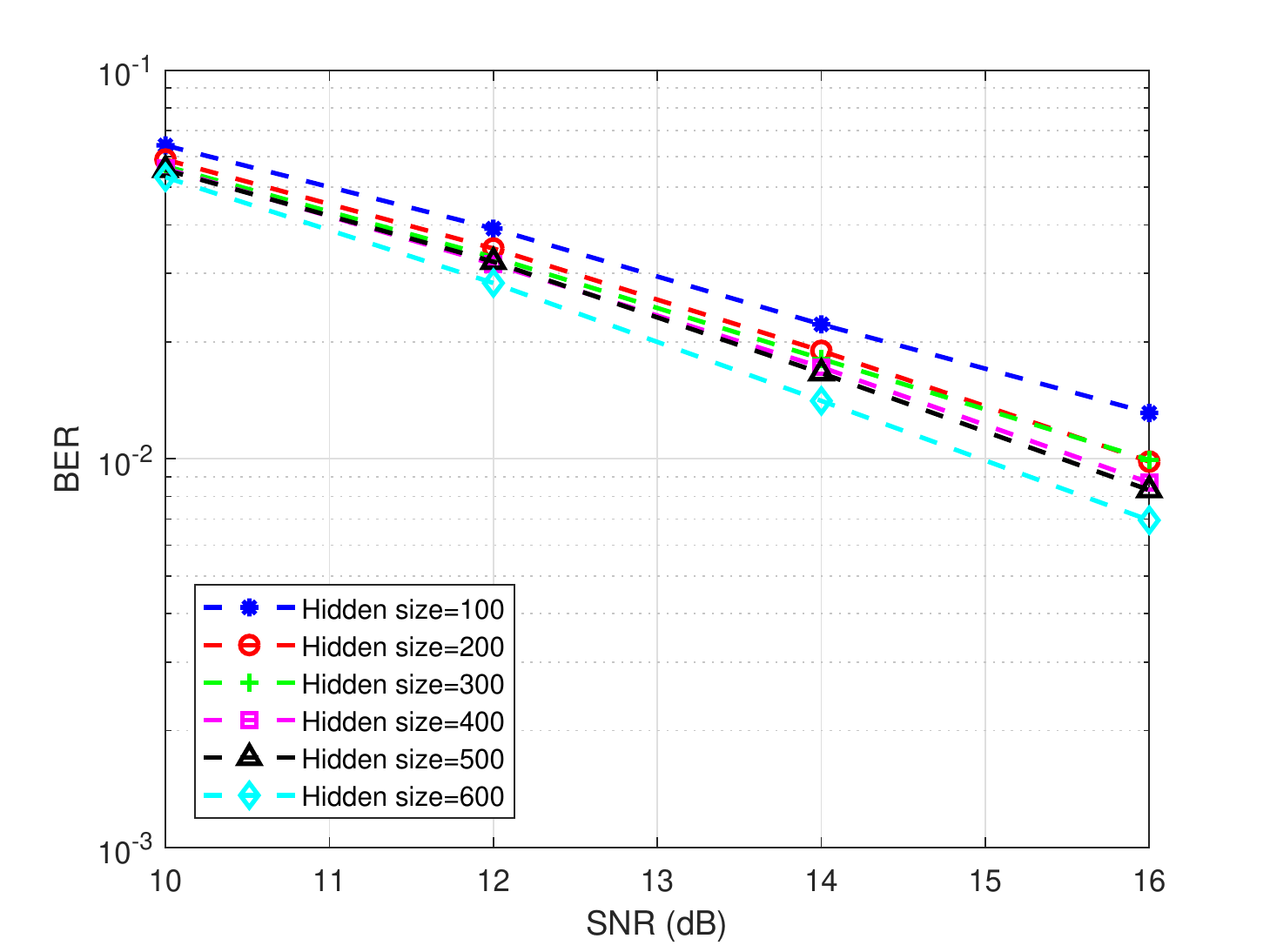}
\caption{The BER performance of LISA with different number of hidden nodes.}
\label{nodes}
\end{figure}

\section{Conclusion}\label{conclusion}

In this paper, we proposed a learning to iterative search approach, named LISA, for the MIMO detection problem. We first proposed a solution construction algorithm using DNN and LSTM as building block for fixed and varying channels, respectively, based on QL-decomposition of the channel matrix. LISA is composed of the unfolding of the solution construction algorithm. Experimental results showed that LISA  is able to achieve near-MLD performance with low complexity both for fixed and varying channel scenarios in QPSK modulation. It performs significantly better than classical detectors and recently proposed machine/deep learning based detectors. Further, we showed that LISA performs significantly better than known detectors for complicated (correlated) channel matrices and imperfect CSI. Extensive experimental results also confirmed that LISA generalises very well against channel correlation and SNR.

\section*{Appendix}

\subsection*{LSTM}

Artificial neural networks (ANN) are computing systems in which nodes are interconnected with each other. The nodes work much like neurons in the human brain. They connect with weights simulating the stimulus among neurons. Together with training algorithms, ANN can recognize hidden patterns and correlations in data. Fig.~\ref{nn} shows a simple feedforward neural network with 3 layers, including an input, a hidden and an output layer. This network receives input and pass it over to the output. A neural network with more than 3 layers is usually called a deep neural network (DNN). A fully connected DNN can be formulated as follows:
\begin{equation}
y = f_m(\cdots f_3(\mathbf{W}_3 f_2(\mathbf{W}_2 f_1(\mathbf{W}_1 + b_1) + b_2) + b_3)\cdots )
\end{equation}where $m$ is the number of layers, $\mathbf{W}_i, b_i$ and $f_i, 1\leq i\leq m$ is the weight matrix, bias and activation function at each layer, respectively.
\begin{figure}[htbp]
\centering
\includegraphics[scale=0.25]{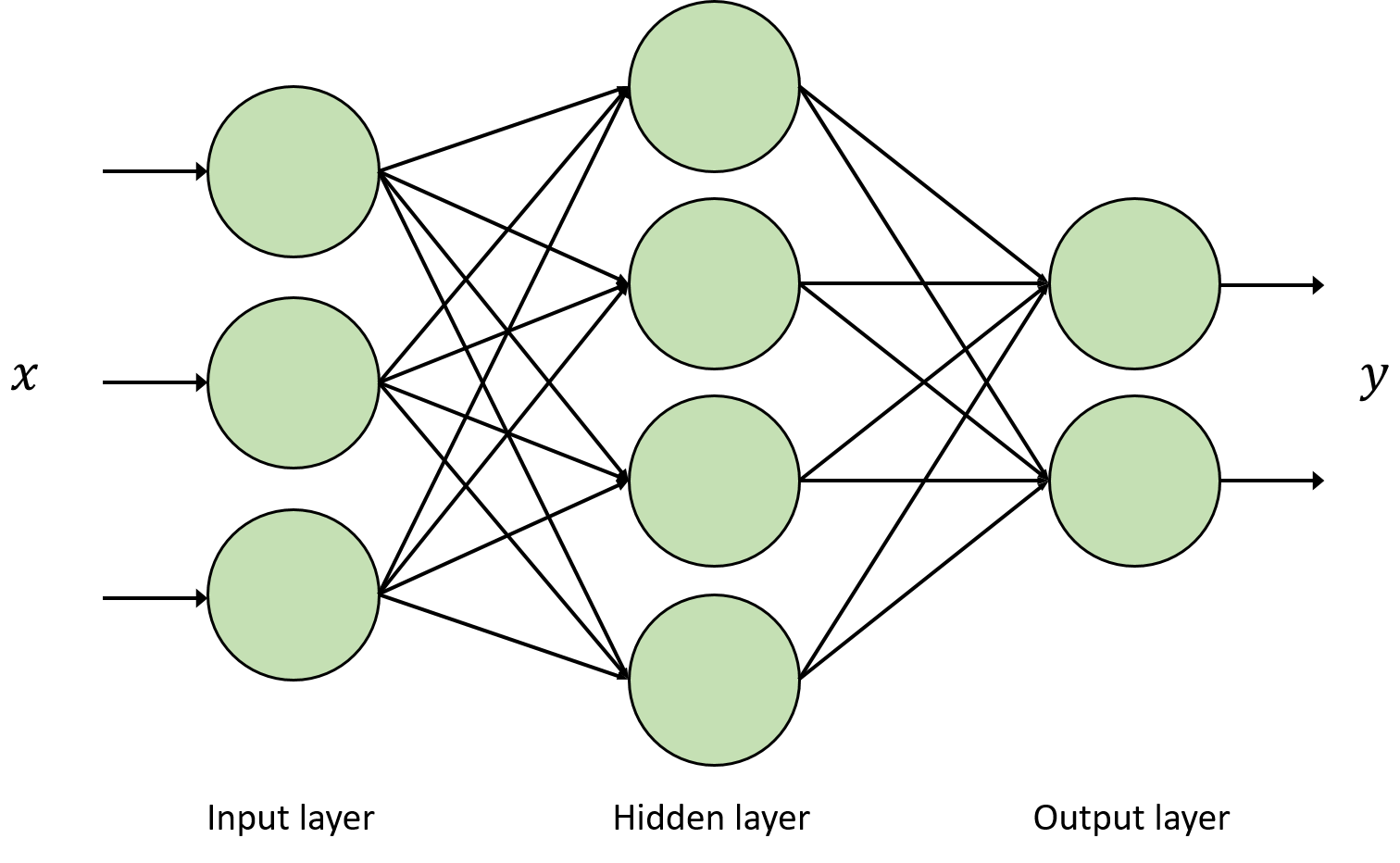}
\caption{A neural network with input, hidden and output layer.}\label{nn}
\end{figure}

ANN cannot process sequential, such as time series, data in which dependencies exist along time. Recurrent neural network (RNN) has been developed to handle this. It is still a kind of neural network but with loop in it. Fig.~\ref{unrolled} shows a recurrent neural network. In the left plot, it is seen that $x_t, t = 0,1, \cdots$ is the input to a neural network (A) and $h_t, t = 0, 1, \cdots$ is the output with a loop in. It can be unrolled as shown in the right plot.~\cite{DBLP:journals/corr/abs-1906-03814}

\begin{figure}[htbp]
\centering
\includegraphics[scale=0.25]{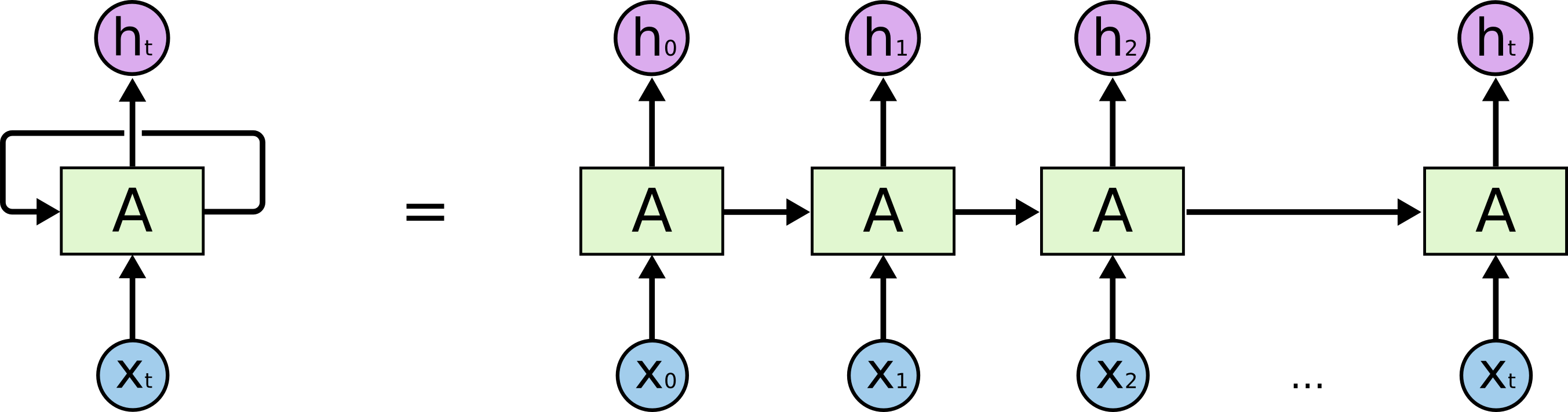}
\caption{An unrolled recurrent neural network where $A$ represents a chunk of neural network, while $x_t, h_t$ is the input and output, respectively. taken from \url{http://colah.github.io/posts/2015-08-Understanding-LSTMs/}.}\label{unrolled}
\end{figure}

Long Short Term Memory networks (LSTM) is a kind of RNN~\cite{Hochreiter1997Long}. It is capable of capturing long-term dependencies among input signals. There are a variety of LSTM variants. In our implementation, we choose the basic LSTM proposed in~\cite{6789445}. The basic LSTM can be formulated as follows:
\begin{eqnarray}
f_t &=& \sigma\left(W_f \cdot [h_{t-1}, x_t] + b_f\right) \nonumber\\
i_t &=& \sigma\left(W_i \cdot [h_{t-1}, x_t]+ b_i\right)  \nonumber\\
\tilde{C}_t &=& \text{tanh}\left(W_C\cdot [h_{t-1}, x_t] + b_C\right) \nonumber\\
C_t &=& f_t \otimes C_{t-1} + i_t \otimes \tilde{C}_t \nonumber\\
o_t &=& \sigma\left(W_o \cdot [h_{t-1}, x_t] + b_o \right) \nonumber\\
h_t &=& o_t \otimes \text{tanh}(C_t) \nonumber
\end{eqnarray}where $[h_{t-1}, x_t]$ means catenation of $h_{t-1}$ and $x_t$, $\otimes$ means element-wise multiplication, $\sigma$ and $\text{tanh}$ is the sigmoid activation function and tanh function, respectively:
\begin{equation}
\sigma(z) = \frac{1}{1+ e^{-z}}; \text{tanh}(z) = \frac{e^z - e^{-z}}{e^z + e^{-z}} \nonumber
\end{equation}It is seen that LSTM is a non-linear function due to the non-linearity of sigmoid and tanh. Parameters of the LSTM include $W_f, b_f, W_i, b_i, W_o$, and  $b_o$. Training algorithms, such as ADAM, are mostly used to find optimal parameters that best fit the training data in the current deep learning community.

\subsection*{ADAM}

The ADAM algorithm~\cite{kingma2014adam} is a type of first-order stochastic gradient descent algorithm based on adaptive estimates of lower-order moments. It computes individual adaptive learning rates for different parameters of a stochastic objective function from estimates of first and second moments of the gradients. ADAM has been widely applied in deep learning for model training. Alg.~\ref{adam} describes the pseudo code of Adam (taken from~\cite{kingma2014adam} with minor modifications).
\begin{algorithm}[h]
 \caption{Adam}\label{adam}
  \KwIn{learning rate $lr$, exponential decay rates for moment estimates $\beta_1, \beta_2 \in [0,1)$ and $f(\bm{\theta})$: stochastic objective function with parameter $\bm{\theta}$}
  initialize $\bm{\theta}_0$
  set $\bm{m}_0 \leftarrow \bm{0}, \bm{v}_0 \leftarrow \bm{0}$ and $t \leftarrow 0$\;
  \While{$\bm{\theta}_t$ not converged}{
  $t \leftarrow t + 1$\;
  $\bm{g}_t \leftarrow \Delta_{\bm{\theta}}f(\bm{\theta}_{t-1})$\;
  $\bm{m}_t \leftarrow \beta_1 \cdot \bm{m}_{t-1} +(1-\beta_1) \cdot \bm{g}_t$\;
  $\bm{v}_t \leftarrow \beta_1 \cdot \bm{v}_{t-1} +(1-\beta_1) \cdot \bm{g}_t^2$\;
  $\hat{\bm{m}}_t \leftarrow \bm{m}_t/(1-\beta_1^t)$\;
  $\hat{\bm{v}}_t \leftarrow \bm{v}_t/(1-\beta_2^t)$\;
  $\bm{\theta}_t \leftarrow \bm{\theta}_{t-1}-lr \cdot \hat{\bm{m}}_t/(\sqrt{\hat{\bm{v}}_t} + \epsilon)$\;
}
\Return $\bm{\theta}_t$
 \end{algorithm}


\bibliographystyle{IEEEtran}
\bibliography{ref}

\begin{thebibliography}{10}
\providecommand{\url}[1]{#1}
\csname url@samestyle\endcsname
\providecommand{\newblock}{\relax}
\providecommand{\bibinfo}[2]{#2}
\providecommand{\BIBentrySTDinterwordspacing}{\spaceskip=0pt\relax}
\providecommand{\BIBentryALTinterwordstretchfactor}{4}
\providecommand{\BIBentryALTinterwordspacing}{\spaceskip=\fontdimen2\font plus
\BIBentryALTinterwordstretchfactor\fontdimen3\font minus
  \fontdimen4\font\relax}
\providecommand{\BIBforeignlanguage}[2]{{%
\expandafter\ifx\csname l@#1\endcsname\relax
\typeout{** WARNING: IEEEtran.bst: No hyphenation pattern has been}%
\typeout{** loaded for the language `#1'. Using the pattern for}%
\typeout{** the default language instead.}%
\else
\language=\csname l@#1\endcsname
\fi
#2}}
\providecommand{\BIBdecl}{\relax}
\BIBdecl

\bibitem{7244171}
S.~Yang and L.~Hanzo, ``Fifty years of {MIMO} detection: The road to
  large-scale {MIMO}s,'' \emph{IEEE Communications Surveys Tutorials}, vol.~17,
  no.~4, pp. 1941--1988, Fourthquarter 2015.

\bibitem{1054829}
G.~Forney, ``Maximum-likelihood sequence estimation of digital sequences in the
  presence of intersymbol interference,'' \emph{IEEE Transactions on
  Information Theory}, vol.~18, no.~3, pp. 363--378, May 1972.

\bibitem{4815548}
E.~G. Larsson, ``{MIMO} detection methods: How they work [lecture notes],''
  \emph{IEEE Signal Processing Magazine}, vol.~26, no.~3, pp. 91--95, May 2009.

\bibitem{zf-df}
G.~Golden, G.~Foschini, R.~Valenzuela, and P.~WOlniansky, ``Detection algorithm
  and initial laboratory results using {V-BLAST} space time communications
  architecture,'' \emph{IEEE Electron. Lett.}, vol.~35, pp. 14--16, 1999.

\bibitem{6375940}
F.~{Rusek}, D.~{Persson}, B.~K. {Lau}, E.~G. {Larsson}, T.~L. {Marzetta},
  O.~{Edfors}, and F.~{Tufvesson}, ``Scaling up {MIMO}: Opportunities and
  challenges with very large arrays,'' \emph{IEEE Signal Processing Magazine},
  vol.~30, no.~1, pp. 40--60, Jan 2013.

\bibitem{7248580}
D.~{Zhu}, B.~{Li}, and P.~{Liang}, ``On the matrix inversion approximation
  based on neumann series in massive {MIMO} systems,'' in \emph{2015 IEEE
  International Conference on Communications (ICC)}, June 2015, pp. 1763--1769.

\bibitem{7401703}
B.~{Kang}, J.~{Yoon}, and J.~{Park}, ``Low complexity massive {MIMO} detection
  architecture based on neumann method,'' in \emph{2015 International SoC
  Design Conference (ISOCC)}, Nov 2015, pp. 293--294.

\bibitem{7370771}
C.~{Tang}, C.~{Liu}, L.~{Yuan}, and Z.~{Xing}, ``High precision low complexity
  matrix inversion based on newton iteration for data detection in the massive
  {MIMO},'' \emph{IEEE Communications Letters}, vol.~20, no.~3, pp. 490--493,
  March 2016.

\bibitem{8638827}
F.~{Jin}, Q.~{Liu}, H.~{Liu}, and P.~{Wu}, ``A low complexity signal detection
  scheme based on improved newton iteration for massive {MIMO} systems,''
  \emph{IEEE Communications Letters}, vol.~23, no.~4, pp. 748--751, April 2019.

\bibitem{6966041}
X.~{Gao}, L.~{Dai}, C.~{Yuen}, and Y.~{Zhang}, ``Low-complexity {MMSE} signal
  detection based on richardson method for large-scale {MIMO} systems,'' in
  \emph{2014 IEEE 80th Vehicular Technology Conference (VTC2014-Fall)}, Sep.
  2014, pp. 1--5.

\bibitem{7342925}
X.~{Qin}, Z.~{Yan}, and G.~{He}, ``A near-optimal detection scheme based on
  joint steepest descent and jacobi method for uplink massive {MIMO} systems,''
  \emph{IEEE Communications Letters}, vol.~20, no.~2, pp. 276--279, Feb 2016.

\bibitem{6954512}
L.~{Dai}, X.~{Gao}, X.~{Su}, S.~{Han}, C.~{I}, and Z.~{Wang}, ``Low-complexity
  soft-output signal detection based on gauss-seidel method for uplink
  multiuser large-scale {MIMO} systems,'' \emph{IEEE Transactions on Vehicular
  Technology}, vol.~64, no.~10, pp. 4839--4845, Oct 2015.

\bibitem{7037314}
X.~{Gao}, L.~{Dai}, Y.~{Hu}, Z.~{Wang}, and Z.~{Wang}, ``Matrix inversion-less
  signal detection using {SOR} method for uplink large-scale {MIMO} systems,''
  in \emph{2014 IEEE Global Communications Conference}, Dec 2014, pp.
  3291--3295.

\bibitem{7399337}
T.~{Xie}, L.~{Dai}, X.~{Gao}, X.~{Dai}, and Y.~{Zhao}, ``Low-complexity
  {SSOR}-based precoding for massive {MIMO} systems,'' \emph{IEEE
  Communications Letters}, vol.~20, no.~4, pp. 744--747, April 2016.

\bibitem{1019833}
E.~Agrell, T.~Eriksson, A.~Vardy, and K.~Zeger, ``Closest point search in
  lattices,'' \emph{IEEE Transactions on Information Theory}, vol.~48, no.~8,
  pp. 2201--2214, Aug 2002.

\bibitem{Damen:2003:MDS:2263399.2271012}
\BIBentryALTinterwordspacing
M.~O. Damen, H.~El~Gamal, and G.~Caire, ``On maximum-likelihood detection and
  the search for the closest lattice point,'' \emph{IEEE Trans. Inf. Theor.},
  vol.~49, no.~10, pp. 2389--2402, Oct. 2003. [Online]. Available:
  \url{https://doi.org/10.1109/TIT.2003.817444}
\BIBentrySTDinterwordspacing

\bibitem{4475373}
J.~Jalden and B.~Ottersten, ``The diversity order of the semidefinite
  relaxation detector,'' \emph{IEEE Transactions on Information Theory},
  vol.~54, no.~4, pp. 1406--1422, April 2008.

\bibitem{7282651}
C.~Jeon, R.~Ghods, A.~Maleki, and C.~Studer, ``Optimality of large {MIMO}
  detection via approximate message passing,'' in \emph{2015 IEEE International
  Symposium on Information Theory (ISIT)}, June 2015, pp. 1227--1231.

\bibitem{6778065}
S.~Wu, L.~Kuang, Z.~Ni, J.~Lu, D.~Huang, and Q.~Guo, ``Low-complexity iterative
  detection for large-scale multiuser {MIMO-OFDM} systems using approximate
  message passing,'' \emph{IEEE Journal of Selected Topics in Signal
  Processing}, vol.~8, no.~5, pp. 902--915, Oct 2014.

\bibitem{mlbook}
T.~M. Mitchell, \emph{Machine Learning}.\hskip 1em plus 0.5em minus 0.4em\relax
  McGraw Hill Education, 2017.

\bibitem{sun19a}
J.~Sun, H.~Zhang, A.~Zhou, Q.~Zhang, K.~Zhang, Z.~Tu, and K.~Ye, ``Learning
  from a stream of non-stationary and dependent data in multiobjective
  evolutionary optimization,'' \emph{IEEE Transactions on Evolutionary
  Computation}, vol.~23, no.~4, pp. 541 -- 555, 2019.

\bibitem{sun19b}
J.~Sun, H.~Zhang, A.~Zhou, Q.~Zhang, and K.~Zhang, ``A new learning-based
  adaptive multi-objective evolutionary algorithm,'' \emph{Swarm and
  Evolutionary Computation}, vol.~44, pp. 304--319, 2019.

\bibitem{shi19}
J.~Shi, Q.~Zhang, and J.~Sun, ``{PPLS/D}: Parallel pareto local search based on
  decomposition,'' \emph{IEEE Transactions on Cybernetics}, vol. Early Access,
  pp. 1--12, 2019.

\bibitem{sun12}
L.~Band, D.~Wells, A.~Larrieu, J.~Sun, A.~Middleton, A.~French, G.~Brunoud,
  E.~Sato, M.~Wilson, B.~Peret, M.~Oliva, R.~Swarup, I.~Sairanen, G.~Parry,
  K.~Ljung, T.~Beeckman, J.~Garibaldi, M.~Estelle, M.~Owen, K.~Vissenberg,
  T.~Hodgman, T.~Pridmore, J.~King, T.~Vernoux, and M.~Bennett, ``Root
  gravitropism is regulated by a transient lateral auxin gradient dependent on
  root angle,'' \emph{PNAS}, vol. 109, no.~12, pp. 4668--4673, 2012.

\bibitem{8422211}
Y.~Huang, P.~P. Liang, Q.~Zhang, and Y.~Liang, ``A machine learning approach to
  {MIMO} communications,'' in \emph{2018 IEEE International Conference on
  Communications (ICC)}, May 2018, pp. 1--6.

\bibitem{8335641}
A.~Elgabli, A.~Elghariani, A.~O. Al-Abbasi, and M.~Bell, ``Two-stage lasso admm
  signal detection algorithm for large scale {MIMO},'' in \emph{2017 51st
  Asilomar Conference on Signals, Systems, and Computers}, Oct 2017, pp.
  1660--1664.

\bibitem{lasso}
R.~Tibshirani, ``Regression shrinkage and selection via the {LASSO},''
  \emph{Journal of the Royal Statistical Society. Series B (Methodological)},
  pp. 267--288, 1996.

\bibitem{admm}
S.~Boyd, N.~Parikh, E.~Chu, B.~Peleato, and J.~Eckstein, ``Distributed
  optimization and statistical learning via the alternating direction method of
  multipliers,'' \emph{Foundations and Trends in Machine Learning}, vol.~3,
  no.~1, pp. 1--122, 2011.

\bibitem{dlbook}
I.~Goodfellow, Y.~Bengio, and A.~Courville, \emph{Deep Learning}.\hskip 1em
  plus 0.5em minus 0.4em\relax MIT Press, 2016.

\bibitem{7966042}
X.~Yan, F.~Long, J.~Wang, N.~Fu, W.~Ou, and B.~Liu, ``Signal detection of
  {MIMO-OFDM} system based on auto encoder and extreme learning machine,'' in
  \emph{2017 International Joint Conference on Neural Networks (IJCNN)}, May
  2017, pp. 1602--1606.

\bibitem{10.1093/nsr/nwx099}
\BIBentryALTinterwordspacing
J.~Sun and Z.~Xu, ``{Model-driven deep-learning},'' \emph{National Science
  Review}, vol.~5, no.~1, pp. 22--24, 08 2017. [Online]. Available:
  \url{https://doi.org/10.1093/nsr/nwx099}
\BIBentrySTDinterwordspacing

\bibitem{lecun}
K.~Gregor and Y.~LeCun, ``Learning fast approximations of sparse coding,'' in
  \emph{Proceedings of the 27th International Conference on Machine Learning},
  2017.

\bibitem{Andrychowicz2016Learning}
M.~Andrychowicz, M.~Denil, S.~Gomez, M.~W. Hoffman, D.~Pfau, T.~Schaul,
  B.~Shillingford, and N.~De~Freitas, ``Learning to learn by gradient descent
  by gradient descent,'' in \emph{Advances in Neural Information Processing
  Systems}, 2016, pp. 3981--3989.

\bibitem{Hochreiter1997Long}
S.~Hochreiter and J.~Schmidhuber, ``Long short-term memory,'' \emph{Neural
  Computation}, vol.~9, no.~8, pp. 1735--1780, 1997.

\bibitem{Li2016Learning}
K.~Li and J.~Malik, ``Learning to optimize,'' \emph{arXiv preprint
  arXiv:1606.01885}, 2016.

\bibitem{Chen2016Learning}
Y.~Chen, M.~W. Hoffman, S.~G. Colmenarejo, M.~Denil, T.~P. Lillicrap, and
  N.~de~Freitas, ``Learning to learn for global optimization of black box
  functions,'' \emph{arXiv preprint arXiv:1611.03824v1}, 2016.

\bibitem{mockus2012bayesian}
J.~Mockus, \emph{Bayesian approach to global optimization: theory and
  applications}.\hskip 1em plus 0.5em minus 0.4em\relax Springer Science \&
  Business Media, 2012.

\bibitem{dai2017}
\BIBentryALTinterwordspacing
H.~Dai, E.~B. Khalil, Y.~Zhang, B.~Dilkina, and L.~Song. (2017, May) Learning
  combinatorial optimization algorithms over graphs. [Online]. Available:
  \url{arXiv:1704.01665v2}
\BIBentrySTDinterwordspacing

\bibitem{8804165}
M.~A. {Albreem}, M.~{Juntti}, and S.~{Shahabuddin}, ``Massive {MIMO} detection
  techniques: A survey,'' \emph{IEEE Communications Surveys Tutorials},
  vol.~21, no.~4, pp. 3109--3132, Fourthquarter 2019.

\bibitem{DBLP:journals/corr/abs-1805-07631}
\BIBentryALTinterwordspacing
N.~Samuel, T.~Diskin, and A.~Wiesel, ``Learning to detect,'' \emph{CoRR}, vol.
  abs/1805.07631, 2018. [Online]. Available:
  \url{http://arxiv.org/abs/1805.07631}
\BIBentrySTDinterwordspacing

\bibitem{EasyChair:376}
G.~Gao, C.~Dong, and K.~Niu, ``Sparsely connected neural network for massive
  {MIMO} detection,'' EasyChair Preprint no. 376, EasyChair, 2018.

\bibitem{DBLP:journals/corr/abs-1812-01571}
\BIBentryALTinterwordspacing
V.~Corlay, J.~J. Boutros, P.~Ciblat, and L.~Brunel, ``Multilevel {MIMO}
  detection with deep learning,'' \emph{CoRR}, vol. abs/1812.01571, 2018.
  [Online]. Available: \url{http://arxiv.org/abs/1812.01571}
\BIBentrySTDinterwordspacing

\bibitem{DBLP:journals/corr/abs-1809-09336}
\BIBentryALTinterwordspacing
H.~He, C.~Wen, S.~Jin, and G.~Y. Li, ``A model-driven deep learning network for
  {MIMO} detection,'' \emph{CoRR}, vol. abs/1809.09336, 2018. [Online].
  Available: \url{http://arxiv.org/abs/1809.09336}
\BIBentrySTDinterwordspacing

\bibitem{DBLP:journals/corr/abs-1907-09439}
\BIBentryALTinterwordspacing
------, ``Model-driven deep learning for joint {MIMO} channel estimation and
  signal detection,'' \emph{CoRR}, vol. abs/1907.09439, 2019. [Online].
  Available: \url{http://arxiv.org/abs/1907.09439}
\BIBentrySTDinterwordspacing

\bibitem{8646369}
X.~{Tan}, Z.~{Zhong}, Z.~{Zhang}, X.~{You}, and C.~{Zhang}, ``Low-complexity
  message passing {MIMO} detection algorithm with deep neural network,'' in
  \emph{2018 IEEE Global Conference on Signal and Information Processing
  (GlobalSIP)}, Nov 2018, pp. 559--563.

\bibitem{6952193}
T.~L. {Narasimhan} and A.~{Chockalingam}, ``Channel hardening-exploiting
  message passing (chemp) receiver in large {MIMO} systems,'' in \emph{2014
  IEEE Wireless Communications and Networking Conference (WCNC)}, April 2014,
  pp. 815--820.

\bibitem{2018arXiv180610827T}
S.~{Takabe}, M.~{Imanishi}, T.~{Wadayama}, and K.~{Hayashi}, ``{Deep
  Learning-Aided Projected Gradient Detector for Massive Overloaded {MIMO}
  Channels},'' \emph{arXiv e-prints}, Jun. 2018.

\bibitem{DBLP:journals/corr/abs-1812-10044}
\BIBentryALTinterwordspacing
S.~Takabe, M.~Imanishi, T.~Wadayama, and K.~Hayashi, ``Trainable projected
  gradient detector for massive overloaded {MIMO} channels: Data-driven tuning
  approach,'' \emph{CoRR}, vol. abs/1812.10044, 2018. [Online]. Available:
  \url{http://arxiv.org/abs/1812.10044}
\BIBentrySTDinterwordspacing

\bibitem{7760475}
R.~{Hayakawa}, K.~{Hayashi}, H.~{Sasahara}, and M.~{Nagahara}, ``Massive
  overloaded {MIMO} signal detection via convex optimization with proximal
  splitting,'' in \emph{2016 24th European Signal Processing Conference
  (EUSIPCO)}, Aug 2016, pp. 1383--1387.

\bibitem{Tan2018Improving}
X.~Tan, W.~Xu, Y.~Be'Ery, Z.~Zhang, X.~You, and C.~Zhang, ``Improving massive
  {MIMO} belief propagation detector with deep neural network,'' \emph{arXiv
  e-prints}, 2018.

\bibitem{7345035}
J.~{Yang}, C.~{Zhang}, X.~{Liang}, S.~{Xu}, and X.~{You}, ``Improved
  symbol-based belief propagation detection for large-scale {MIMO},'' in
  \emph{2015 IEEE Workshop on Signal Processing Systems (SiPS)}, Oct 2015, pp.
  1--6.

\bibitem{DBLP:journals/corr/abs-1906-03814}
\BIBentryALTinterwordspacing
Y.~Wei, M.~Zhao, M.~Zhao, and M.~Lei, ``Learned conjugate gradient descent
  network for massive {MIMO} detection,'' \emph{CoRR}, vol. abs/1906.03814,
  2019. [Online]. Available: \url{http://arxiv.org/abs/1906.03814}
\BIBentrySTDinterwordspacing

\bibitem{Shirkoohi2019AdaptiveNS}
M.~K. Shirkoohi, M.~Alizadeh, J.~Hoydis, and P.~Fleming, ``Adaptive neural
  signal detection for massive {MIMO},'' \emph{ArXiv}, vol. abs/1906.04610,
  2019.

\bibitem{fcsd}
L.~Barbero and J.~Thompson, ``Fixing the complexity of the sphere decoder for
  {{MIMO}} detection,'' \emph{IEEE Trans. Wireless Commun.}, vol.~7, pp.
  2131--2142, 2008.

\bibitem{6789445}
F.~A. {Gers}, J.~{Schmidhuber}, and F.~{Cummins}, ``Learning to forget:
  Continual prediction with {LSTM},'' \emph{Neural Computation}, vol.~12,
  no.~10, pp. 2451--2471, Oct 2000.

\bibitem{1408197}
J.~{Jalden} and B.~{Ottersten}, ``On the complexity of sphere decoding in
  digital communications,'' \emph{IEEE Transactions on Signal Processing},
  vol.~53, no.~4, pp. 1474--1484, April 2005.

\bibitem{1468474}
B.~{Hassibi} and H.~{Vikalo}, ``On the sphere-decoding algorithm {I}. expected
  complexity,'' \emph{IEEE Transactions on Signal Processing}, vol.~53, no.~8,
  pp. 2806--2818, Aug 2005.

\bibitem{kingma2014adam}
D.~P. Kingma and J.~Ba, ``Adam: A method for stochastic optimization,''
  \emph{arXiv preprint arXiv:1412.6980}, 2014.

\bibitem{8227772}
N.~{Samuel}, T.~{Diskin}, and A.~{Wiesel}, ``Deep {MIMO} detection,'' in
  \emph{2017 IEEE 18th International Workshop on Signal Processing Advances in
  Wireless Communications (SPAWC)}, July 2017, pp. 1--5.

\bibitem{5773010}
A.~{Ghasemmehdi} and E.~{Agrell}, ``Faster recursions in sphere decoding,''
  \emph{IEEE Transactions on Information Theory}, vol.~57, no.~6, pp.
  3530--3536, June 2011.

\bibitem{1683382}
C.~{Oestges}, ``Validity of the kronecker model for {MIMO} correlated
  channels,'' in \emph{2006 IEEE 63rd Vehicular Technology Conference}, vol.~6,
  May 2006, pp. 2818--2822.

\bibitem{8398483}
F.~{Jiang}, C.~{Li}, and Z.~{Gong}, ``Accurate analytical {BER} performance for
  {ZF} receivers under imperfect channel in low-snr region for large receiving
  antennas,'' \emph{IEEE Signal Processing Letters}, vol.~25, no.~8, pp.
  1246--1250, 2018.

\end{thebibliography}
\end{document}